\documentclass{jpp}
\usepackage[]{graphicx}
\usepackage{booktabs}
\usepackage{epstopdf, epsfig}
\usepackage{hyperref}
\usepackage{changepage}
\usepackage{xspace}
\usepackage{natbib}
\usepackage{amsmath}
\usepackage{xcolor}

\allowdisplaybreaks

\newcommand{\graphsubdir}[1]{#1}
\newcommand{\figref}[1]{Fig.~\ref{fig:#1}}
\newcommand{\appref}[1]{Appendix~\ref{app:#1}}
\newcommand{\tabref}[1]{Tab.~\ref{tab:#1}}
\newcommand{\coderunner}{\textsc{CodeRunner}\xspace}
\newcommand{\trinity}{\textsc{Trinity}\xspace}
\newcommand{\corfu}{\textsc{CoRFu}\xspace}
\newcommand{\gryfx}{\textsc{Gryfx}\xspace}
\newcommand{\gstwo}{\textsc{Gs2}\xspace}
\newcommand{\chease}{\textsc{Chease}\xspace}

\newcommand{\gyroR}[1]{\left< #1 \right>_{\mathbf{R}}}
\newcommand{\gyror}[1]{\left< #1 \right>_{\mathbf{r}}}
\newcommand{\vct}[1]{\mathbf{#1}}
\newcommand{\uv}[1]{\hat{\mathbf{#1}}}
\newcommand{\NL}[1]{\mathcal{N}_{#1}}
\newcommand{\pderiv}[2]{\frac{\partial #1}{\partial #2}}
\newcommand{\sgam}{\Gamma_0^{1/2}}
\newcommand{\vPsi}{{\bf v}_\Phi}
\newcommand{\vE}{{\bf v}_E}
\newcommand{\flr}{\frac{1}{2} \hat{\nabla}_\perp^2}

\newcommand{\fsa}[1]{\left< #1 \right>_\psi}

\shorttitle{Optimisation of fusion confinement}
\shortauthor{E. G. Highcock, N. R. Mandell, M. Barnes and W. Dorland}

\title{Optimisation of confinement in a fusion reactor using a nonlinear turbulence model}

\author{E. G. Highcock\aff{1,2,3}
\corresp{\email{highcock@chalmers.se}},
N. R. Mandell\aff{4},
M. Barnes\aff{2},
\and W. Dorland\aff{5}}

\affiliation{\aff{1}Department of Physics, Chalmers University of Technology, Gothenburg, Sweden
\aff{2}Rudolph Peierls Centre for Theoretical Physics, 1 Keble Road, Oxford, UK
\aff{3}Culham Centre for Fusion Energy, Culham Science Centre, Abingdon, UK
\aff{4}Department of Astrophysical Sciences, Princeton University, Princeton, New Jersey, USA
\aff{5}Department of Physics, University of Maryland, College Park, Maryland, USA
}

\begin{document}

\maketitle

\begin{abstract}
  The confinement of heat in the core of a magnetic fusion reactor is optimised using a multidimensional optimisation algorithm.
  For the first time in such a study, the loss of heat due to turbulence is modelled at every stage using first-principles nonlinear simulations which accurately capture the turbulent cascade and large-scale zonal flows.
  The simulations utilise a novel approach, with gyrofluid treatment of the small-scale drift waves and gyrokinetic treatment of the large-scale zonal flows.
  A simple near-circular equilibrium with standard parameters is chosen as the initial condition.
  The figure of merit, fusion power per unit volume, is calculated, and then two control parameters, the elongation and triangularity of the outer flux surface, are varied, with the algorithm seeking to optimise the chosen figure of merit.
  A two-fold increase in the plasma power per unit volume is achieved by moving to higher elongation and strongly negative triangularity.
\end{abstract}

\section{Introduction}
Convective heat loss resulting from micro-turbulent fluctuations in a fusion reactor limits the ability of such a reactor to confine heat to the degree required to achieve fusion.
Above some critical temperature gradient, this heat loss rises extremely rapidly,
limiting the central temperature and hence the fusion reaction rate.
%in many cases effectively ``pinning'' the temperature gradient close to its critical value; this phenomenon is known as stiff transport (of which there are several reviews, e.g. \cite{Tynan2009,Doyle2007,Horton1999a}).
%With the temperature gradient limited,
Empirically, one way to improve fusion performance despite this obstacle is to build larger devices.%
\footnote{This is based on the observation that the global confinement of heat improves with size~\citep{Doyle2007}, which is
believed to stem from the corresponding increase in the height of the spontaneously-formed transport barrier
which forms at the edge of the plasma in typical conditions (see e.g.~\citealp{Maggi2007a}).}
Current planning for the first demonstration power plant \citep{Federici2014}
envisions a device more than 20 times larger by volume than the world's largest operating test reactor, the Joint European Torus (JET).
Such a tactic remains the surest route to achieving fusion with our current understanding.
However, the approach is not without difficulties. First, the individual cost of such a large plant is very high (although the power output scales accordingly), putting the construction of such reactors, as well as necessary precursor experiments and test reactors, out of the reach of all but major governments.
Second, the larger designs place much greater demands upon the construction materials (see e.g.\ the discussion in~\cite{Stork2014}).
%The enormous magnetic fluxes need advanced high-strength steels to support them;
%since the power produced increases more rapidly than the area available for the power to flow out,
%plasma facing materials need to be able to withstand correspondingly higher thermal and neutron loads.
%since the power produced increases more rapidly than the area available for the power
%to flow out the thermal and neutron loads needing to be withstood are
%correspondingly higher.
%since power produced scales with the volume of the device but the area
%available for the power to flow out scales with the radius,
%the thermal and neutron loads needing to be withstood are
%correspondingly higher.

The reactor design effort is constantly finding innovative ways to mitigate these problems.
%: to reduce the demands on materials, to make it easier to replace them, and so on.
For example, one approach to achieving more compact and cost-effective designs is to use new high-temperature superconductors which have lower cost and greater tensile strength than the superconductors used today, and which can be constructed so as to allow easy disassembly for maintenance \citep{Sorbom2015}.

However, while such approaches have frequently used state of the art modelling for almost all of the construction of the reactor, none of these studies have incorporated the turbulent heat loss using nonlinear models which accurately capture the essential properties of the turbulence.
Instead, the vast majority use scaling laws, either inspired by dimensional arguments or extrapolated from current experiments \citep{Uckan1990,Galambos1995,Luce2014,Sorbom2015} (see also discussion in~\cite{Zohm2013}).
It is also possible to use more sophisticated ``quasilinear'' models \citep{Staebler2007, Bourdelle2016} which calculate the linear properties of the instabilities which drive the turbulence and use phenomenological rules, and calibration to a pre-determined subset of nonlinear simulations, to calculate the turbulent heat loss from these linear properties.
Use of these quasilinear models includes both full reactor design studies (e.g.\ \cite{Staebler2006, Wenninger2015, Jardin2006}) and
efforts to predict and optimise performance for particular devices (e.g.\ \cite{Mukhovatov2003,Budny2009,Kinsey2011,Parail2013,Meneghini2016}).
Additionally, there have been advances in using neural-network approaches to allow such methods to fully replace the simple scaling laws by facilitating extremely fast simulation based on the quasilinear models \citep{Citrin2015a}.
Nonetheless, nonlinear turbulence calculations, while used routinely for investigation of individual experiments and verification of quasilinear models, have been conspicuously absent from reactor design.%
\footnote{In stating this we have not overlooked the important work across the entire field 
(which a lack of space prevents us from surveying)
studying the performance of and possible improvements to \textit{existing} designs such
as ITER and DEMO using nonlinear simulations. 
It is the success of such work that motivates inclusion of nonlinear analysis
in the design process itself.}

The primary reason for the absence of nonlinear models is their cost and complexity.
%of which more will be described later.
%The primary reason for this is their cost and complexity.
Effectively, it would have been impossible until recently to have completed a design study including these models on any reasonable timescale.
This absence represents a missed opportunity.
%because simplified models of turbulent heat loss are used, a conscientious fusion designer must use pessimistic estimates in order to allow margin for error (see discussion in \cite{Zohm2013}).
Over the last 20 years there has been a great increase in understanding, derived from both experiment and theoretical inquiry, of the circumstances in which the levels of turbulence can be greatly reduced, without reducing the temperature gradient \citep{Burrell1997,Synakowski1999,Highcock2010,Barnes2011,Highcock2012,Citrin2015}.
Such effects can be, and have been, included in quasilinear and other models \textit{a posteriori}.
However, even where designs are based on extrapolation from the best performing experimental configurations, using the most complete reduced models available at the time (e.g.~\cite{Jardin2006}), the designs can know nothing of potential new phenomena, leading to possible further gains, that might exist in the vast configuration space at their disposal
(more than two decades ago,~\cite{Galambos1995} showed that in theory, where full control could be maintained over turbulent transport, it was possible to reduce the capital cost of the reactor plant by 30\%, that is, 10 billion (1995) US dollars, compared to the most advanced designs of the day).
This still-largely-uncharted configuration space is best explored in conjunction with nonlinear models, since any exploration by reduced models could conceivably miss a crucial physics effect not yet included within the model%
\footnote{As discussed in more detail below, this proof-of-principle study includes only an electrostatic, adiabatic-electron, hybrid gyrofluid-gyrokinetic nonlinear model, but the methodology demonstrated is already capable of being extended to include a full gyrokinetic electromagnetic multi-species calculation, albeit at increased cost.}
(c.f\ work in the stellarator community which clearly demonstrates the benefit of tight integration of stellarator optimisation with nonlinear analysis;~\cite{Xanthopoulos2014}).

Every fusion experiment costs a vast sum to build, and owing to the hard limits placed by all components of the reactor, can only explore a certain small region of the high- (or infinite-) dimensional design space.
It is beholden upon theory and numerical study to search the design space, at several-orders-of-magnitude smaller expense, to identify the most promising regions.
However, in order for this search to be meaningful, we must have confidence in the predictions.
The simplified models used must either be pessimistic, or be used only in regimes where they have been benchmarked against more complete models or experiment and thus preclude the possibility of entirely new design paradigms.
By contrast, careful comparison with experiment over recent years has provided a high degree of confidence that nonlinear models (specifically gyrokinetic models, which use a five-dimensional reduction of the Vlasov equation valid in the conditions of a fusion reactor; for a review see~\cite{Abel2013}), without any provision of tuning or fitting parameters, are able to accurately predict what the properties of turbulence will be like in a given situation~\citep{White2013,Citrin2014,VanWyk2016}.

\section{A first-principles model}

Here we present the first case where a first-principles nonlinear model of turbulence (specifically a novel hybrid gyrofluid/gyrokinetic model, described below, which produces excellent agreement with gyrokinetic models) has been used, as part of a multi-scale transport analysis, to calculate the performance of the core of a given configuration \textit{ab initio} and then seek for a higher performing solution.
In this particular study, the optimisation algorithm chosen achieved an improvement in the fusion power per unit volume of 91\%.
However, this first study is envisioned as a proof of concept, as the stepping stone for a much larger effort in which many more dimensions of parameter space are explored.

\begin{figure}
  \begin{center}
    \includegraphics[width=0.65\textwidth]{\graphsubdir{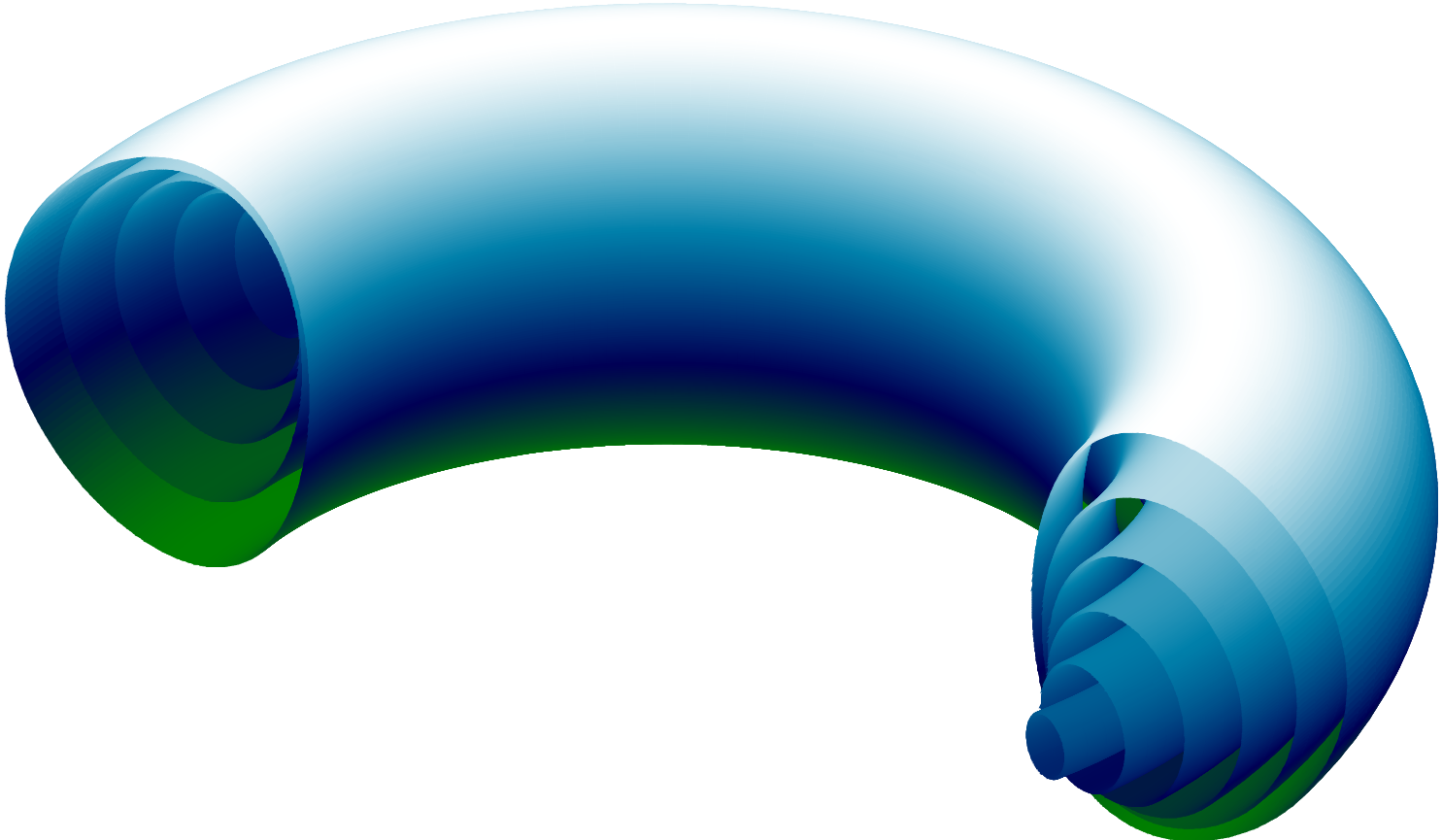}}
  \end{center}
  \caption{An illustration of the nested magnetic flux surfaces that confine a fusion plasma. The shape of these fluxes corresponds to the initial shape used for this study.}\label{fig:nestedfluxsurfaces}
\end{figure}

A magnetic fusion reactor is composed of a plasma confined by a series of nested toroidal magnetic flux surfaces (\figref{nestedfluxsurfaces}).
The plasma is composed of ionised deuterium and tritium, which fuse together to produce helium, neutrons, and a large amount of energy, provided sufficient pressure is achieved over sufficient volume in the centre of the plasma.
The magnetic field is provided by a series of large external field coils and a current which flows through the plasma itself (as well as by various small coils responsible in addition for the stability of the field).
The plasma is heated in a number of different ways: by the current itself, by the injection of a beam of high energy neutral particles, and by high power electromagnetic radiation of various frequencies. Fuel is injected by means of puffs of gas, frozen pellets, and neutral particle beams.
Heat is lost via neutrons, radiation and by particle loss to the wall (which is concentrated in a special target region called the divertor).
In this investigation we consider the core of the reactor; that is we consider the inner $\sim$90\% (by minor radius) of the flux surfaces.
We assume  the existence of an edge transport barrier, that is, that there is good confinement with a steep pressure gradient in the remaining $\sim$10\% of the plasma
(a standard assumption since the discovery of the ``high-confinement mode'';~\cite{Wagner1984}).
We also assume that the barrier is independent of elongation and triangularity (this assumption, necessary because we are considering only core transport, differs from current experimental observations, and is discussed further below).

\begin{figure}
  %\begin{raggedleft}center
  %\begin{center}
  \centerline{%
  %\begin{adjustwidth}{}{1cm}
  \includegraphics[width=0.65\textwidth]{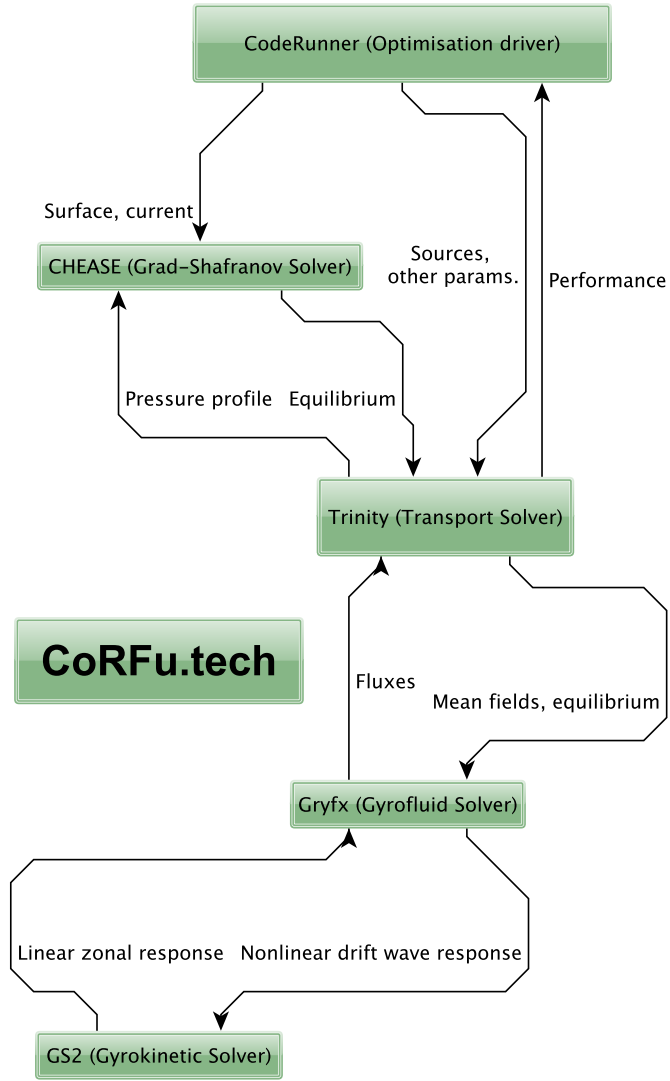}\hspace*{2em}
  %\end{adjustwidth}
  }
  %\end{center}
  %\end{raggedleft}
  \caption{Schematic of the \corfu framework.}\label{fig:schematic}
\end{figure}

The software used in this study to optimise such a reactor is the \coderunner Fusion Optimisation Framework (\corfu).
This is a sizeable edifice of software comprising many independent parts, as illustrated in \figref{schematic}.
The equations underlying the framework are given in Appendices~\ref{app:equations}--\ref{app:zonal}.
The calculation proceeds as follows.

To begin with, the choice of optimisation control parameters, and all other requisite input information,
is provided to the \corfu framework, which acts as the optimisation driver.
From this input, an initial configuration is assembled, comprising parameterisations of:
\begin{itemize}
  \item the shape of the flux surface at the edge of the plasma core;
  \item the profile of the toroidal current, and
  \item the initial pressure profile, including a fixed finite pressure at the edge of the plasma core.
\end{itemize}
These are used (Link 1 in \figref{schematic}) by the \chease code \citep{Bondeson1996} to calculate the shape of the magnetic flux surfaces (i.e., the equilibrium magnetic field).

\begin{figure}
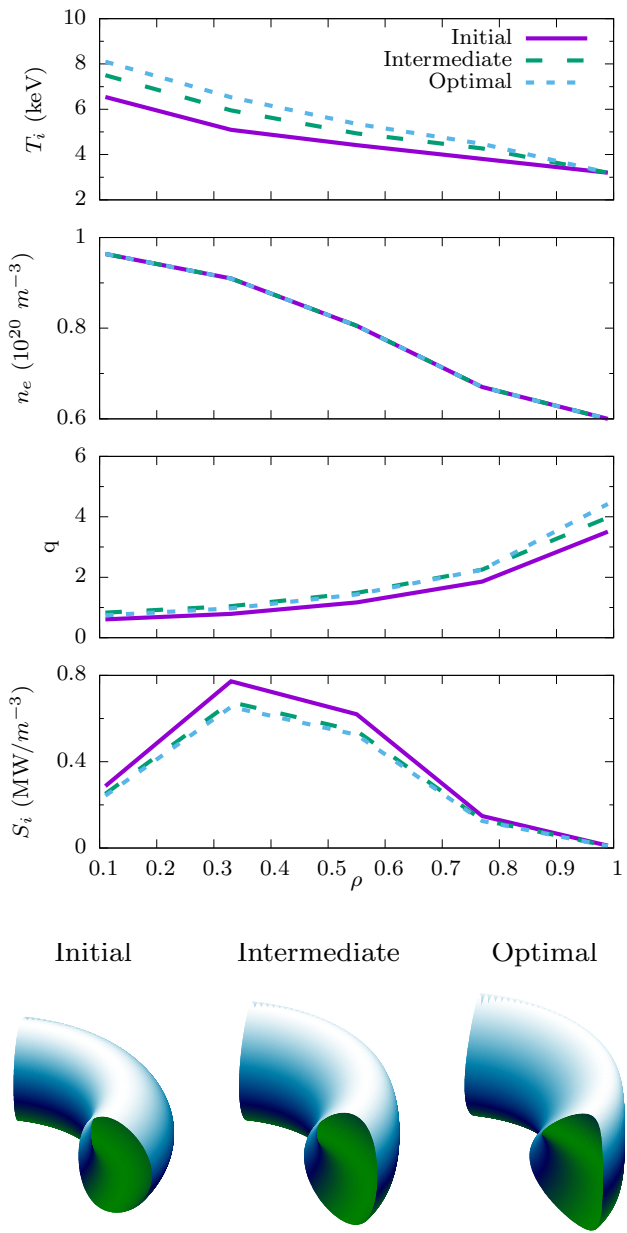

  \begin{center}
    \includegraphics[width=0.60\textwidth]{\graphsubdir{initialfinal}}
    \vspace*{2em}

    \includegraphics[width=0.60\textwidth]{\graphsubdir{elongtriangcartoon}}
  \end{center}
  \caption{\textit{Top}. Profiles of ion temperature ($T_i$), electron density ($n_e$), safety factor ($q$) and heat input ($S_i$), showing the values of each quantity versus normalised radius $\rho$.
  %in both the starting (green, solid) and optimal (blue, dashed) configurations.
  %Note that fusion power is a sensitive function of ion temperature, and that the small changes in $T_i$ nonetheless produce significant changes in fusion power.
  $\rho$ is a dimensionless quantity which labels the flux surface. In this work we examine core transport; thus at the outer flux surface ($\rho=1$) there is a finite temperature of 3.2 keV and a finite electron density of 0.6$\times10^{20}$~m$^{-3}$.
  \textit{Bottom}. Outer flux surface shapes of the initial, intermediate and optimal solutions.
  }\label{fig:initialfinal}
\end{figure}

%The system of equations in this study produces a self-consistent transport solution regardless of the exact choice of density and source profiles (rotation is neglected in this study) , which are specified using simple paramterisations.
%The initial configuration is chosen to be comparable in magnitude (though by no means exactly the same as) a real experimental situation: this allows us to assess whether the solution is roughly credible \textit{a posteriori}. Machine parameters are based on the Joint European Torus (JET).
For the initial configuration we consider a \emph{purely hypothetical} tokamak.
We do, however, choose machine parameters (see Table~\ref{tab:globalresults}) that are comparable in magnitude to those of a large tokamak---the Joint European Torus (JET)---and profiles that are plausible for such a tokamak (\figref{initialfinal}; see e.g.~\cite{Barnes2011} or the tokamak profile database described in~\cite{Roach2008} for a comparison).

%and
%choose density and source profiles (both held fixed here) based on those of JET discharge \#42982, as treated in \citep{Barnes2010} and distributed in the tokamak profile database \citep{Roach2008}.
%These profiles are displayed in \figref{trinitycomp}.
%While the system of equations used in this study produces a self-consistent result without a need for these particular choices, choosing a configuration comparable in magnitude to a real experiment allows us to assess the credibility of the initial solution, by comparing the predicted temperature profile of the initial solution with that measured for discharge \#42982: see \figref{trinitycomp}.

%\begin{figure}
%\begin{center}
%\includegraphics[width=0.5\textwidth]{\graphsubdir{trinity_comp.eps}}
%\end{center}
%\caption{Although this study considers a hypothetical configuration, we choose the electron density ($n_e$), ion heat source ($S_i$) and safety factor ($q$) profiles (here we show the initial case) to be the same order of magnitude as a real experiment (JET shot \#42982).
%The calculated ion temperature ($T_i$) profile for the initial case agrees with the experimental case to within an order of magnitude.}
%\label{fig:trinitycomp}
%\end{figure}

The initial configuration, including the magnetic equilibrium solution, details of external heat sources and the profiles of particle density (densities and external sources are held fixed) are then used (Links 2 and 3) by the \trinity transport solver \citep{Barnes2010}
to find the pressure (and thus the fusion power generated), across the whole of the plasma, with the additional assumption that the temperatures of the ions
and electrons are kept equal by collisional equilibration.\footnote{This limits the possibility of interplay between the electron and ion heating and loss channels, either to good or ill; we note again that this and other limitations of this proof-of-principles study are discussed more fully below.}
This is done by allowing the pressure profile to evolve until the heat losses,
due to turbulence (discussed below) and other effects (in this study, neoclassical transport and radiative losses, see~\appref{equations}),
match the heat inputs, both external and that generated by fusion.
However, since the magnetic equilibrium itself is affected by the pressure, the result of the \trinity calculation must be fed back to \chease (Link 4), a new equilibrium generated, and \trinity rerun, until the cycle converges, the pressure stops changing and a steady-state solution is found.

\begin{figure*}
  \begin{center}
    \includegraphics[width=\textwidth]{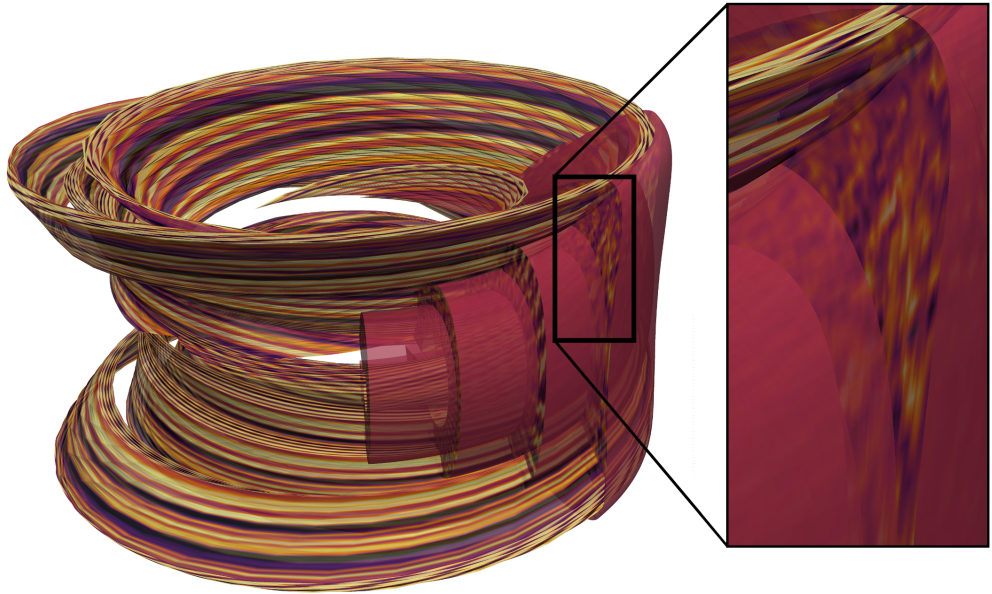}
  \end{center}
  \caption{An illustration of a \trinity simulation showing above-threshold turbulence for the optimal configuration. The elongated
  ribbons to the left of the main picture are flux tubes \citep{Beer1995}, the domains in which the turbulence is calculated,
  using the local approximation which assumes the turbulent fluctuations are small-scale perpendicular to the magnetic field.
  Assuming statistical periodicity (which again results from the small scale of the fluctuations), these flux tubes can be
  repeated to fill the four flux surfaces being simulated (which are shown cut away to the right of the main picture). The
  picture to the right shows the cross-section of the turbulence.
  %Its anisotropic nature (small-scale across the field lines,
  %elongated along the field lines) is apparent by comparing this cross-section to the structure along the flux-tubes (which
  %follow the magnetic field lines).
  }\label{fig:infernocompiledzonal}
\end{figure*}

To calculate the heat loss due to turbulence (Links 5 and 8),
\trinity uses several simultaneous copies of the GPU-based
hybrid gyrofluid/gyrokinetic code \gryfx (see Appendices~\ref{app:gyrofluid}--\ref{app:zonal}).
Each copy of \gryfx calculates
%a separate element of the Jacobian for the implicit evolution of the temperature
%by calculating
the heat flux resulting from the turbulence at a particular location in the plasma, that is,
on a particular magnetic flux surface, given the
pressure profile and magnetic equilibrium.
A mean-field, multi-scale approach is used. The turbulent heat flux
is calculated for the current pressure profile; the heat flux is then used
to evolve the pressure profile, at which point the turbulent heat flux
is recalculated.
An illustration of the complete system is given in \figref{infernocompiledzonal}.

%with the appropriate temperature gradient perturbation.
The \gryfx code divides the calculation of the heat flux in two (Links 6 and 7) using a novel algorithm:
the evolution of the smaller scale drift waves is calculated using the gyrofluid model \citep{Dorland1993a,Beer1996a} by \gryfx,
and the evolution of the large-scale zonal flows which are generated by the turbulence is calculated by the gyrokinetic code \gstwo \citep{Dorland2000,Kotschenreuther1995c},
with the nonlinear interaction between the two scales (and among the drift waves of various scales) being calculated in \gryfx (see Appendix~\ref{app:zonal}).
The simulations are electrostatic with a Boltzmann response for electrons.
%Every 1000 \gryfx timesteps \trinity pauses \gryfx and determines whether the turbulence calculation has reached saturation, in other words, when it has converged upon a value of the heat flux.
%(this is a step fraught with challenges, as discussed below).
%When each copy of \gryfx has reached a saturated (steady) state, it passes the heat flux back to \trinity (Link 8), which uses the fluxes to evolve the pressure profile, and then calls \gryfx again with the new pressure profile. This cycle continues until the pressure profile has adjusted so that the heat loss owing to turbulence matches the heat injection, and the pressure profile stops changing.

%At this point, \trinity exits, and passes the pressure profile to \chease which updates the magnetic equilibrium. \trinity is then run again. This loop is then repeated until both the pressure profile and the magnetic equilibrium stop changing; in other words, a steady state solution has been calculated.

%\section{Choice of initial configuration}

%This study seeks to show how any generic configuration can be
%The transport calculation within \trinity used only five radial grid points.
%This was to reduce the overall cost of this first study, and would be increased in future work.
%Nonetheless, we note that,
%owing to the 5-point derivative stencil used radially by \trinity
%and the fact that the transport calculations evolved to steady state
%(where power balance was satisfied on each flux surface),
%it is likely that the sparse radial grid would not have had a large impact on the results of this study.
The transport calculation within \trinity used only five radial grid points.
This was to reduce the overall cost of this first study, and would be increased in future work.
Nonetheless, we note that the 5-point derivative stencil used radially by Trinity,
and the fact that the transport calculations evolved to steady state (where power balance was satisfied on each flux surface),
will reduce the impact of the sparse radial grid on the results of this study.

\section{Finding the optimal solution}

\begin{figure}
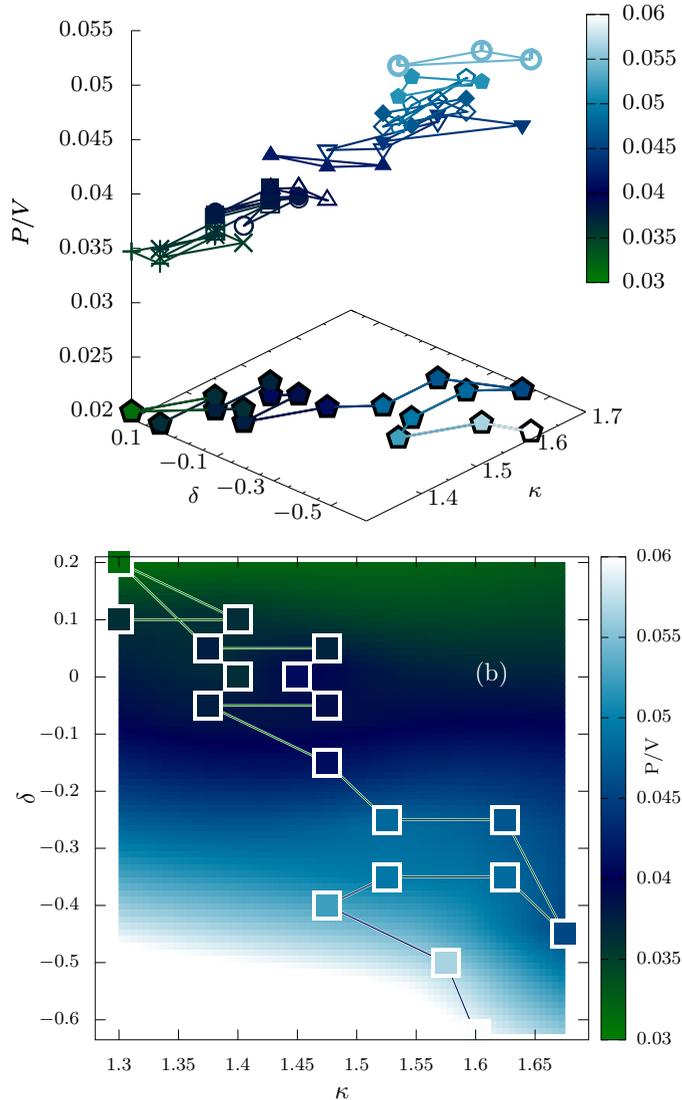

  \begin{center}
    \includegraphics[width=0.65\textwidth]{\graphsubdir{simplex}}
    \vspace{1em}

    \includegraphics[width=0.65\textwidth]{\graphsubdir{interp}}
  \end{center}
  \caption{\textit{Top}. Diagram illustrating the search path taken by the simplex optimisation algorithm.
  The suspended triangles represent the simplex at each iteration, with the colour and height of each triangle equal to the average of the values of $P/V$ at each of the vertices. The curve in the base plane indicates the actual sequence of function calls, that is, the actual sequence of ($\kappa$, $\delta$) values evaluated by \trinity.
  \textit{Bottom}. $P/V$ as a function of $\kappa$ and $\delta$, both the function values (squares) and an interpolated surface, with the colour being the value of $P/V$.
  The highest (optimal) value is at $\kappa=1.60$, $\delta=-0.625$. An intermediate solution, with a triangularity within the bounds of
  what has currently been realised, is chosen at $\kappa=1.525$, $\delta=-0.25$.
  %both the function values (squares) and an interpolated surface. The results indicate firstly an optimum triangularity of around 0.05 and secondly that, at this triangularity, an elongation of around 1.55 has optimum performance.
  }\label{fig:simplex}
\end{figure}

When the pressure, magnetic equilibrium, and turbulent heat fluxes have ceased to evolve, a steady-state solution has been found.
At this point, the figure of merit for this solution (the fusion power per unit volume) is used by the optimisation driver \coderunner (Link 9 in \figref{schematic}), to generate a new set of control parameters according to the particular optimisation algorithm being used.
In this study, the Nelder-Mead simplex algorithm was selected\footnote{The simplex algorithm
was chosen for its reliability in this initial study.
However, the \corfu framework is expected to be used to explore a multi-dimensional design space
with a plethora of local maxima.
This next step will first require a careful study of the suitability of various
multi-dimensional optimisation algorithms both standard and novel, whether heuristics like simulated annealing or genetic algorithms,
and whether gradient-free like Nelder-Mead, or pseudo-gradient like BFGS and variants.
\cite{Lewis2004b} provides a good overview. %chktex 2
The field of nonlinear multi-dimensional optimisation is large and increasing,
with both commercial~\citep{modefrontier} and freely available~\cite{Abramson2006} software offerings.
Wide, too, is the literature on the comparison of algorithms, whether individual
examples \citep{Manousopoulos2009} or more abstract papers investigating the techniques
of algorithm comparison, or metamodelling \citep{Simpson2008}.}
(not to be confused with the simplex algorithm in linear programming).
This is a slow but robust algorithm that constructs a simplex with N+1 vertices in an N-dimensional configuration space and then at each stage takes the ``worst'' vertex and moves it to a more optimal place.
Visually, one can perceive the simplex ``crawling uphill'' (\figref{simplex}).

Using the new values of the control parameters provided by the simplex algorithm (i.e.\ the next ``guess'') the whole of the previous cycle is repeated, determining the figure of merit for this next configuration.
The calculation proceeds likewise until it has converged on a maximum to within some specified tolerance (or until a prescribed number of iterations have taken place).

%There are many choices of algorithms with various pros and cons, but

%\begin{figure}[htb]
%\begin{center}

%\includegraphics[width=0.29\textwidth]{\graphsubdir{elongtriang}}
%\end{center}
%\caption{
%\textit{Top}. In the current study, the control parameters were the elongation ($\kappa$) and the triangularity ($\delta$) of the outer flux surface (that is, flux surface at the edge of the core). \textit{Bottom}. The outer flux surface is plotted for the first three pairs of these control parameters considered, and also for the optimal solution (gold).
%}
%\label{fig:elongtriang}
%\end{figure}

The two control parameters chosen were the elongation and triangularity of the outer flux surface (as defined in~\cite{Bondeson1996}). The shape of the magnetic field has long been known to have a marked effect on the turbulence, and the choice of these two shaping parameters in particular was motivated by pioneering experimental work on the TCV tokamak, which was designed to allow large variation in both---demonstrating a consequent large variation in performance~\citep{Weisen1999}.

%Typically the resolution of a turbulent calculation is adjusted manually on a case by case basis (after observing intermediate results), but in the present instance the resolution (and domain size) of the turbulence calculation must be set to be large enough to handle all phenomena that may be encountered.
%The secondary challenge is that each simulation must be run until it has converged on a steady-state (non-pathological) solution. Once again, in the typical case a researcher can manually observe when a simulation has converged and detect any pathologies.
%In this case the \trinity code has to heuristically determine whether or not a flux solution is unconverged or invalid.

The results are displayed at the bottom of \figref{simplex}. In this figure one can observe the evolution of the control parameters, starting with three initial guesses of ($\kappa$, $\delta$) = (1.3, 0.2), (1.3, 0.1) and (1.4, 0.1). The lowest initial vertex is (1.3,0.2) with a power per unit volume ($P/V$) of 0.0313~MW\,m$^{-3}$. Initially the algorithm moves to higher elongation and lower triangularity, before proceeding to keep elongation roughly constant and to continue to decrease triangularity. The iteration was terminated when values of triangularity moved significantly beyond the limits of what has been seen in experiment ($\sim$-0.65~\cite{Pochelon1999}). The final, optimal value of  $P/V$ of 0.0598~MW\,m$^{-3}$ with ($\kappa$, $\delta$) = (1.6, -0.625).
Thus, an improvement of 91\% was discovered over the course of the optimisation. Since the final value of triangularity was somewhat extreme, we also provide data for an intermediate result with ($\kappa$, $\delta$) = (1.525, -0.25), for which $P/V$ is 0.0481~MW\,m$^{-3}$.

\begin{figure}
  %\begin{center}
  \centerline{\includegraphics[width=0.65\textwidth]{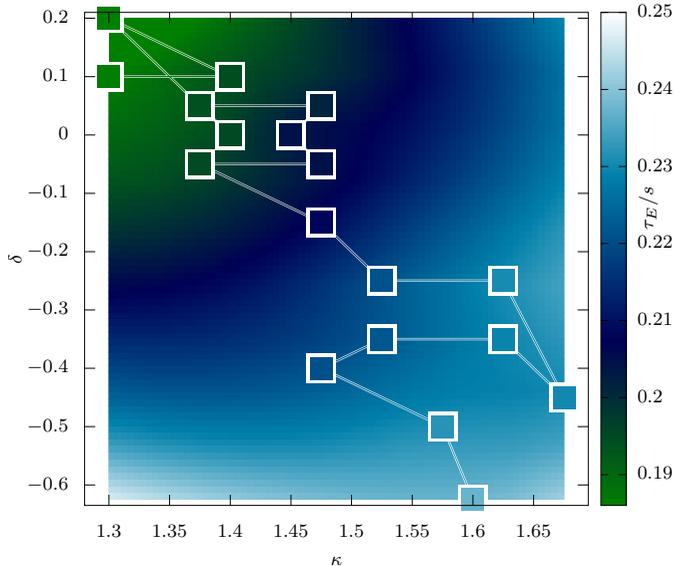}}
  %\end{center}
  \caption{Evolution of the confinement time in seconds, given as a function of
  elongation $\kappa$ and triangularity $\delta$. Squares represent data points,
  shown on top of an interpolated surface for illustration of the overall trend.}\label{fig:confinementtime}
\end{figure}

It is interesting also to consider the evolution of the confinement time, defined as the total amount of stored kinetic energy divided by the
rate of energy injection, is displayed in \figref{confinementtime}.
%We find that it is the same order of magnitude
%as that determined for JET \#42982, given in the profile database \cite{Roach2008} as 0.4s (the differences between the exact details
%of this study and the quoted case, as well as the uncertainties in determining the result, explain the discrepancy).
We also see that optimising the confinement time, which reflects the ability of the system to confine heat, would have produced different results to optimising the fusion power per unit volume; in particular, the improvement at negative triangularity is less marked, and moving to higher elongation at positive triangularity may have produced a similar improvement in the confinement time.

\section{Discussion and Outlook}

\begin{table*}
  %\small
  \centering
  \begin{tabular}{rccc}
    & Initial & Intermediate & Optimal \\
    \toprule
    On-axis Magnetic Field. (T) & 2.91 & 2.84 & 2.81 \\
    On-axis Major Radius (m) & 3.12 & 3.17 & 3.21 \\
    Ion input heat $S_i$ (MW) & 20 & 20 & 20 \\
    Plasma current $I_p$ (MA) & 1.9 & 1.9 & 1.9\\
    Elongation $\kappa$  & 1.3 & 1.525 & 1.60 \\
    Triangularity $\delta$  & 0.2 & -0.25 & -0.625 \\
    Core Volume $V$ (m$^3$) & 59.3 & 67.1 & 71.8\\
    Confinement time (s) & 0.179 & 0.221 &  0.237\\
    Fusion Power $P$ (MW) & 1.85 & 3.23 & 4.30 \\
    $P/V$ (MW/m$^3$) & 0.0313 & 0.0481 &  0.0598\\
    \bottomrule
  \end{tabular}
  \caption{Global properties of the initial, intermediate and optimal configurations}\label{tab:globalresults}
\end{table*}

It has been shown that, without any tuning of model parameters, it has been possible to generate a plausible solution for reactor performance and then optimise it.
The final fusion power in this study is small compared to the input heating power (on the ions) of 20MW, but that is because the parameters in this study (\tabref{globalresults}) are not reactor-like, but rather of the same order of magnitude as those of JET, the largest existing device: a choice made to allow assessment of the credibility of the initial solution.
%Turning to the profiles of various important quantities in \figref{initialfinal}, it can be seen, by comparison with \cite{Barnes2010}, that with sensible choices for the density, heat deposition and safety factor profiles
%the initial temperature profile is entirely plausible
%in this theoretical study we have merely constructed an initial setup which is JET-like and then allowed the optimisation algorithm to search in any direction.
Even more encouragingly, the qualitative result of this optimisation---that core confinement improved at lower triangularity and higher elongation---is directly corroborated by experimental measurements on the TCV tokamak~\citep{Weisen1999} and theoretical studies~\citep{Marinoni2009}, although as TCV is a much smaller device than that we consider here, with important effects arising from kinetic electron modes and finite gyroradius, quantitative comparisons cannot be justified.
%Maybe plot JET42982?

The challenges inherent to this study, and the obstacles to such a study being carried out in the past, proceed principally from the calculation of the turbulent fluxes.
%The primary challenge is the expense of such a calculation.
All in all a total of 8680 converged nonlinear turbulence calculations at a spatial resolution of 192$\times$96$\times$20 (radial$\times$poloidal$\times$parallel) were required, at a total cost of approximately 3000 GPU-accelerated node hours.
That such an exercise was possible was owing to two developments. The first was the design of the implicit algorithm in the \trinity code which reduced the number of turbulence calculations required~\citep{Barnes2010}.
%The second was the development of the hybrid GPU-based gyrofluid/gyrokinetic code \gryfx,
The second was the development of \gryfx,
which by using a gyrokinetic response for the zonal flows and advanced nonlinear closures was able to overcome the shortcomings of gyrofluid codes in the 1990s~\citep{Dimits2000} while still maintaining their huge speed advantage over gyrokinetic codes (typically a GRYFX run will be 20 times faster and 100 times less expensive than, for example, a \gstwo run;~\cite{Mandell2014}).

It is only right to point out in this discussion that this study focuses solely on the plasma core, and not on the reactor components or the plasma edge.
It should be noted that though edge transport barriers can be achieved in combination with negative triangularity~\citep{Pochelon2012}, the evolution of the temperature at the edge of the domain, here held constant, would be strongly altered if the edge of the plasma was modelled in addition to the core, and thus the problem of optimizing the performance of the plasma as whole is distinct from optimizing that of the core, as is discussed further in~\citep{Pochelon2012}. In fact it is likely that the edge confinement would degrade as the triangularity becomes negative, reducing the favourability of negative triangularity regimes (see e.g.~\cite{Merle2017}).
Thus further work is needed to understand how these results would be modified if the plasma edge was modelled self-consistently with the core.
In light of these limitations, it would be most desirable in the future to incorporate this methodology in a holistic model of the device (including checks on magneto-hydrodynamic stability);
in particular it would be desirable to control the safety factor profile to maintain values greater than one in the core to avoid the sawtooth instability (the small increase in the safety factor required is not expected to significantly affect the results presented here).
%, and to include
%the effects of rotation, kinetic electrons and so on.
However, while ultimately all parts of a tokamak must be considered, by finding ways to improve core performance, the demands placed upon the reactor design and the requirement for high plasma confinement in the edge can be lessened.
%It is legitimate to seek for configurations with exceptional core performance (respecting known engineering limits), and then work backwards to determine how such a configuration can be achieved in practice.

Having demonstrated the first successful optimisation of (core) confinement using a nonlinear model of turbulence, there are several directions that are immediately attractive to follow. The first is to use this technique to find optimal configurations for JET ahead of its first run using an active (deuterium-tritium) fuel mix in twenty years (using parameters that match JET far more closely than in this study). The second is to consider, in a similar fashion, ways in which the performance of ITER, the new, global fusion experiment being constructed, can be improved. It would also be important to extend \gryfx to consider electromagnetic effects and kinetic electron physics, and to include the effects of rotation, which are not considered here.
Other priorities would be to include additional transport channels: electron and ion heat and particles, momentum, impurity transport and so on.
The complex interplay between these channels can be the crucial factor in achieving higher performance (see e.g.~\cite{Highcock2011}), and robust operation, and is an ideal area to study with a nonlinear framework such as \corfu.
The eventual goal is to switch to using optimisation algorithms which parallelize easily and can search for global maxima, and run a vast parallel blue-skies search for a dramatically optimised fusion reactor.

\section{Acknowledgements}

The authors of this paper are indebted to A. Schekochihin, T. F\"ul\"op, G. Hammett, J. Ball, J. Citrin, M. Landremann, I. Abel, F. van Wyk, F. Parra, C. Roach,  S. Newton, I. Pusztai and J. Omotani for helpful discussions, ideas and support.
This work has been carried out within the framework of the EUROfusion Consortium and has received funding from the Euratom research and training programme 2014--2018 under grant agreement No 633053.
The views and opinions expressed herein do not necessarily reflect those of the European Commission.
This work has also been supported by the Framework grant for Strategic Energy Research (Dnr. 2014--5392) from Vetenskapsr\aa{}det. 
NRM is supported by the U.S. Department of Energy via the DOE CSGF program, provided under grant DE-FG02-97ER25308. %chktex 8
This work used the TACC Stampede supercomputer and the ARCHER UK National Supercomputing Service (http://www.archer.ac.uk).

%We are grateful for illuminating discussions with A. Schekochihin, T. F\"ul\"op, J. Citrin, M. Landremann and I. Abel.
%Simulations were carried out using the TACC Stampede supercomputer and the ARCHER UK National Supercomputing Service (http://www.archer.ac.uk).
%This work has been carried out within the framework of the EUROfusion Consortium and was supported by a EUROfusion fusion researcher fellowship [WP14-FRF-CCFE/Highcock]. The views and opinions expressed herein do not necessarily reflect those of the European Commission.

%\section{Author contributions}
%
%E.H. performed the simulations, developed the CodeRunner optimisation framework and co-developed \trinity, \gstwo and \gryfx. N.M. was the primary developer of the hybrid gyrofluid-gyrokinetic code \gryfx and provided support for running the simulations.
%M.B. developed \trinity both originally and additionally for this study, including the important extension to handle multiple species, and co-developed \gstwo.
%W.D. co-developed \gryfx (including the theoretical framework) and \trinity, developed \gstwo and the support for general geometry within both \gstwo and \gryfx, and provided advice and support for the project as a whole.

%\end{refsection}
\appendix

%\begin{refsection}
\section{Equations underlying the optimisation framework}\label{app:equations}

The equations that govern the system are presented in (\cite{Abel2013}), which builds upon an already large body of work (significantly~\cite{Frieman1982,Sugama1998}).

They follow a multi-scale approach, with a clear separation in time
between the evolution of the safety factor (the twist of the magnetic field lines, which evolves on the resistive timescale;~\cite{Abel2013a}),
the evolution of the pressure (which evolves on the confinement timescale), and the evolution of the turbulence (which occurs on the timescale of the linear micro-instabilities which drive it). The equations governing all three are
\begin{equation}
  \frac{\partial q}{\partial t} = \frac{c}{4\pi^2}\frac{\partial }{\partial \psi} V' \left< \vct{E}\cdot \vct{B} \right>_{\psi}
  \label{eqn:safetyfactor}
\end{equation}
\begin{equation}
  \frac{3}{2}\frac{1}{V'}\frac{\partial }{\partial t} V'\left< n_s \right>_{\psi}T_s +
  \frac{1}{V'} \frac{\partial }{\partial \psi}  V' \left< Q_s \right>_{\psi}
  = S_s
  \label{eqn:pressureevolution}
\end{equation}
and
\small
\begin{multline}
  %\left( \frac{\partial }{\partial t} + \vct{u}\cdot \nabla \right)h_s +\\
  \frac{\partial h_s}{\partial t} +
  \left(v_{\parallel} \uv{b} + \vct{V}_{D_s} +
  \frac{c}{B}\uv{b}\times \nabla\gyroR{\varphi}  \right) \cdot \nabla h_s \\
  = \gyroR{C\left[ h_s \right] } + \frac{Z_se F_s}{T_s}\frac{\partial \gyroR{\varphi}}{\partial t} - %chktex 25
  \frac{\partial F_s}{\partial \psi}  \left( \uv{b} \times \nabla \gyroR{\varphi} \right)\cdot \nabla\psi,
  \label{eqn:gk}
\end{multline}
\normalsize
respectively.
In these equations, $q$ is the magnetic safety factor, $\psi$ is the poloidal magnetic flux which is contained within a flux surface, $V$ is the volume of the flux surface, $V'=dV/d\psi$ is the incremental volume element (loosely, the flux surface area), $\vct{E}$ is the electric field and $\vct{B}$ the magnetic field, and $\left<  \right>_\psi$ denotes an average over the flux surface. The index $s$ labels the charged particle species (e.g.\ electron, deuterium ion, tritium ion etc.), $T_s$ and $n_s$ are the species temperature and density, $Q_s$ is the flux of heat across a flux surface, and $S_s$ is a volumetric heat source.
The quantity $h_s$ is the fluctuating (turbulent) part of the particle distribution function, $v_{\parallel}$ is the velocity along the field line, $\uv{b}=\vct{B}/\left| \vct{B} \right|$, $\vct{V}_{D_s}$  represents the effect of the magnetic field inhomogeneities on particle motion, $\varphi$ is the perturbed electric potential, $C$ represents the effects of collisions between particles, $Z_se$ is the species charge and $F_s$ is the equilibrium (slowly varying) component of the species distribution function. %chktex 25
The operator $\gyroR{}$ denotes an average over gyrophase at fixed gyrocentre location $\vct{R}$.
The turbulent component of the heat flux $Q_s$ can be trivially obtained from the turbulent distribution function $h_s$  
\begin{equation}
  Q_s = \left< \int d^{3}\vct{v}\frac{m_s v^{2}}{2}\left[ \left( \frac{c}{B}\uv{b}\times \nabla\gyroR{\varphi} \right)\cdot\nabla \psi \right]\gyror{h_s} \right>_{\psi,t}
  \label{eq:Qs}
\end{equation}
where $\gyror{}$ denotes a gyroaverage at constant particle position.
The (usually) sub-dominant neoclassical component of the heat flux is modelled using analytical approximations~\citep{Chang1982}, as are radiative losses due to Bremsstrahlung~\citep{Glasstone1960}.

Since we are seeking a steady state solution, %with a prescribed current profile,
%\eqref{eqn:safetyfactor} is not required for this study.
equation~\eqref{eqn:safetyfactor} is not evolved during this study.
Instead we  prescribe a fixed profile of surface averaged toroidal current.
This causes $q$ to change with the pressure gradient, not on the resistive timescale, during the evolution towards steady state, but once steady state is reached $q$ ceases to evolve and the solution is self-consistent.
Using the prescribed profile of toroidal current, a prescribed outer flux surface and the pressure profile, the poloidal flux (and hence the shape of the magnetic flux surfaces) can be determined using the Grad-Shafranov equation:
\begin{equation}
  \Delta^* \psi =
  -4\pi R^2 \sum_s n_s T_s\left\{ \frac{d \ln n_s}{d\psi} + \frac{d\ln T_s}{d\psi}\right\}
  - I(\psi)\frac{dI}{d\psi},
\end{equation}
where $\Delta^*$ is the Grad-Shafranov operator
\begin{equation}
  \Delta^* \psi =\left( \frac{\partial ^2}{\partial R^2}  - \frac{1}{R} \frac{\partial }{\partial R} + \frac{\partial ^2}{\partial Z^2} \right) \psi.
\end{equation}
Note that when using the \chease code (\cite{Bondeson1996}) to solve the Grad-Shafranov equation, the usual function $I\partial I / d \psi$ can be replaced as an input by the surface averaged toroidal current density, which is what is specified in this study.

The equation for the pressure (\ref{eqn:pressureevolution}) is evolved by \trinity (\cite{Barnes2010}; \trinity is also capable of evolving the density and rotation, which are kept fixed for this study).
This equation determines how the evolution of the pressure is governed by the total flux of heat, given sources and the shape of the magnetic flux surfaces (that is, the solution of the Grad-Shafranov equation).
%(it is the change of the flux surface area which results from the change in pressure, as the magnetic surfaces expand and contract, which results in the need to repeatedly call \trinity followed by \chease until the shape of the flux surfaces converges, as described in the main body).
Since the solution of the Grad-Shafranov equation itself varies on the same timescale as the pressure, the solution is periodically updated with the new pressure profile until steady-state is reached.
The pressure equation can be evolved separately for each charged species; in this study, to reduce expense, it is solved for the deuterium ions. It is then assumed that collisional processes will rapidly equilibrate the temperatures of all species, and hence that the temperatures of the electrons and tritium ions is the same as that of the deuterium ions.
A fixed pressure is set at the outer magnetic flux surface ($\rho=1$); this is set to a pressure to be expected at the edge of the core, that is, at the top of an edge transport barrier.

%In order to minimise the number of flux evaluations required, \trinity uses an implicit algorithm to increase the size of the time steps taken in evolving equation \eqref{eqn:pressureevolution}.
%As the equation is fully non-linear, with the fluxes depending in an opaque way upon the profiles, \trinity uses a Newton algorithm to search for the future profiles. While leading to big gains in speed, the Newton algorithm is very sensitive to errors in the value of the turbulent fluxes, and consequently requires extra care in the handling of the code which calculates them.

The turbulent distribution function, from which can be calculated the turbulent fluxes, is determined by the gyrokinetic equation~\eqref{eqn:gk}. This can be rigorously derived from first principles using assumptions which are valid in a fusion reactor. However, it remains too expensive to solve for the purposes of this study.

\begin{figure}
  \begin{center}
    \includegraphics[width=0.65\textwidth]{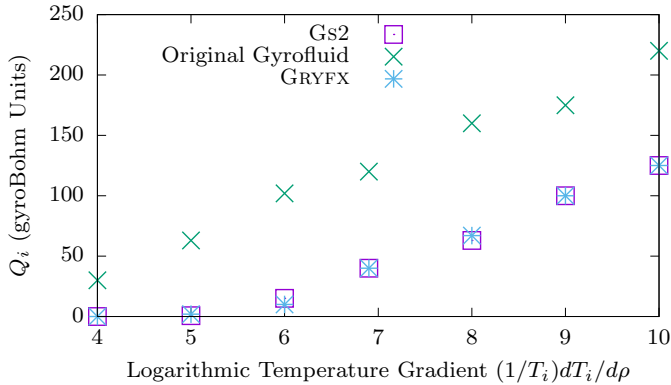}
  \end{center}
  \caption{Comparison between gyrokinetic simulations (\gstwo),
  the original gyrofluid model (\cite{Beer1996a}), and \gryfx. The turbulent heat flux
  is shown as a function of the ion temperature gradient for Cyclone Base Case \citep{Dimits2000} parameters.}\label{fig:gryfxgs2comp}
\end{figure}

Instead, velocity space integrals are taken of this equation to create a hierarchy of fluid moments.
In the early 1990s, a set of closures for this hierarchy were developed which accurately captured the linear response for drift waves, as well as finite Larmor radius effects~\citep{Beer1996a, Dorland1993a, Hammett1990}.
Unfortunately, while successful in many cases, these ``gyrofluid'' models were unable to capture correctly two key properties of the turbulence, namely, the excitation of large-scale zonal flows~\citep{Dimits2000} and the phenomenon of perpendicular nonlinear phase mixing~\citep{Tatsuno2009}.
This resulted in the over-prediction of heat fluxes by gyrofluid models.
However, in recent years a new hybrid gyrofluid/gyrokinetic code has been developed, \gryfx (see Appendices~\ref{app:gyrofluid}--\ref{app:zonal}),
which overcomes these weaknesses, producing excellent agreement with codes that
solve the gyrokinetic equation (\figref{gryfxgs2comp}) whilst still taking orders of magnitude less
time.
As is described in the main text, a principle advance has been the use of a gyrokinetic solver for the linear zonal flow response (Appendix~\ref{app:zonal}). 
This enables \gryfx to capture the characteristic ``Dimits shift''~\citep{Dimits2000}, where, for the parameters used in \figref{gryfxgs2comp}, the turbulent transport is suppressed to negligible levels in the range $4\leq(1/T_i)dT_i/d\rho \lesssim 5$ even though the temperature gradient is above the linear stability threshold. This nonlinear upshift in the threshold has long been attributed to zonal flow dynamics, and our results are consistent with this theory.
At large temperature gradients, the turbulence is stronger and the zonal flows are not as effective at suppressing the turbulence. In this regime, new closures that model the effects of nonlinear phase mixing (Appendix~\ref{app:NLPM}) are the dominant effect in reducing the gyrofluid heat flux predictions to gyrokinetic levels.
It is also important to note that recent theoretical and numerical work has showed that higher
velocity space moments are energetically sub-dominant in the regimes of
interest, producing renewed confidence that a closure with a sufficient number
of moments can capture the important dynamics~\citep{Schekochihin2016,Parker2016a}.  In addition since \gryfx
includes the full quadratic nonlinearity (the fourth term on the left hand side of equation~\eqref{eqn:gk}), it is expected to capture important new
phenomena such as subcritical turbulence.

When using \gryfx to calculate the turbulent fluxes within \trinity, it is essential to use sufficient resolution to resolve the turbulent phenomena, to check for any pathologies, and to ensure that the turbulent calculation has reached convergence. There are many turbulent phenomena that can make this challenging, particularly near the threshold for the onset of turbulence, including long timescale oscillations and large amplitudes of the zonal flows.
This is covered further in \appref{practicalconsiderations}.

%It is also important, for the Newton iteration, to ensure that the flux calculation is converged to a degree smaller than the change between one iteration and the next.

\section{Gyrofluid equations}\label{app:gyrofluid}

The gyrofluid model in \gryfx is based on the 4+2 toroidal gyrofluid model of~\citep{Beer1996a}, which describes the time evolution of six guiding center moments of the gyrokinetic equation,~\eqref{eqn:gk}:
density ($n$), parallel velocity ($u_\parallel$), parallel and perpendicular temperature ($T_\parallel$ and $T_\perp$), and the parallel fluxes of parallel and perpendicular heat ($q_\parallel$ and $q_\perp$). 
These moments can be defined via the following velocity space integrals of the gyroaveraged fluctuating distribution function $g_s = h_s - (Z_se/T_s)\gyroR{\varphi}F_s$:%chktex 25
\begin{equation*}
  \begin{gathered}
    \delta n = \smallint g\ d^3v \\
    \delta p_\parallel = T_s\delta n + n_s\delta T_\parallel = m_s \smallint g\ v_\parallel^2\ d^3v \\
    \delta q_\parallel = -3 n_s T_s \delta u_\parallel + m_s\smallint g\ v_\parallel^3\ d^3v \\
  \end{gathered}
  \qquad \quad
  \begin{gathered}
    n_0 \delta u_\parallel = \smallint g\ v_\parallel \ d^3 v \\
    \delta p_\perp= T_s\delta n + n_s\delta T_\perp =  (m_s/2) \smallint g\  v_\perp^2\ d^3v \\
    \delta q_\perp = -n_s T_s \delta u_\parallel + (m_s/2) \smallint g\  v_\parallel v_\perp^2 \ d^3 v
  \end{gathered}
\end{equation*}
These fluctuating quantities (along with the electrostatic potential $\varphi$) are normalized as
\begin{align}
  &\frac{a}{\rho_i} \left( \frac{\delta n}{n_s},\ \frac{\delta u_\parallel}{v_{t,s}},\ \frac{\delta T}{T_s},\ \frac{\delta q}{n_s m_s v_{t,s}^3},\ \frac{e \varphi}{T_i}\right) = \left(\tilde{n},\ \tilde{u}_\parallel,\ \tilde{T},\ \tilde{q},\ \tilde{\varphi} \right)
\end{align}
where $v_{t,s} = \sqrt{T_s/m_s}$ is the ion thermal speed, $n_s$ and $T_s$ are the equilibrium ion density and temperature respectively, $\rho_s$ is the ion gyroradius, and $a$ is the normalizing length defined to be half the diameter of the last closed flux surface at the midplane. Here the subscript $i$ denotes the reference ion species.
These moment definitions follow~\citep{Snyder2001} and are consistent with Beer's original  definitions. Note that we will evolve $\tilde{T}_{\parallel}$ and $\tilde{T}_\perp$, whereas Beer used $\tilde{p}_{\parallel}=\tilde{n}+\tilde{T}_\parallel$ and $\tilde{p}_\perp=\tilde{n}+\tilde{T}_\perp$. Hereafter quantities will be normalised and non-dimensionalised unless otherwise specified, so we will drop the tildes unless they are necessary for clarity.\footnote{Our non-dimensionalisation is chosen to be consistent with Appendix A of~\citep{Mandell2018}.}
Further, note that these normalised moments are equivalent to the first several Laguerre-Hermite velocity moments of $g$, as described in~\citep{Mandell2018}:
\begin{align}
  \left( G_{0,0},\ G_{0,1},\ \sqrt{2} G_{0,2} ,\ \sqrt{6} G_{0,3},\ G_{1,0},\ G_{1,1} \right) =
  \left(n,\ u_\parallel,\ T_\parallel,\ q_\parallel,\ T_\perp,\ q_\perp
  \right), \label{Beermoments}
\end{align}
where the Laguerre-Hermite moments are defined by (again in non-dimensional form)
\begin{equation}
  G_{\ell,m} = 2\pi \int_{-\infty}^\infty d v_\parallel
  \int_{0}^\infty d\mu B\ \frac{(-1)^\ell}{\sqrt{m!}} \mathrm{L}_\ell(\mu B) \mathrm{He}_m(v_\parallel) g \label{eq:project} %chktex 3
\end{equation}
with $\mathrm{L}_\ell$ the Laguerre polynomials and $\mathrm{He}_m$ the probabilist's Hermite polynomials.

The gyrofluid equations that we solve in \gryfx can then be written as
\begin{align}
  \pderiv{n}{t}
  &+\NL{n} + B \nabla_\parallel \frac{u_\parallel}{B}- \left( \frac{1}{L_{ns}} + \frac{1}{2 L_{Ts}} \hat\nabla_\perp^2 \right) i\omega_\star \Phi \notag\\
  &+ \left( 2 + {1 \over 2} {\hat \nabla}_{\perp}^2 \right) i \omega_d \Phi + i \omega_d \left( T_\parallel + T_\perp + 2n \right) = 0  \label{eq:gf1}\\
  \pderiv{u_\parallel}{t}
  &+ \NL{u_\parallel}+ B \nabla_\parallel \frac{n+T_\parallel}{B} + \nabla_\parallel \Phi  + \left( T_\perp + n + {1 \over 2} {\hat \nabla}_{\perp}^2\Phi \right) \nabla_\parallel \ln B\notag \\
  & + i \omega_d \left( q_\parallel + q_\perp + 4u_\parallel \right) = 0  \\
  \pderiv{T_\parallel}{t}
  &+ \NL{T_\parallel} + B \nabla_\parallel \frac{q_\parallel + 2u_\parallel}{B} + 2 \left( q_\perp + u_\parallel \right) \nabla_\parallel \ln B - \frac{1}{L_{Ts}} i \omega_\star \Phi + 2 i \omega_d \Phi \notag\\
  &+ i \omega_d \left( 6 T_\parallel + 2n \right) + 2 \left| \omega_d \right| \left( \nu_1 T_\parallel + \nu_2 T_\perp \right) = - \frac{2}{3} \nu_{ss} \left( T_\parallel - T_\perp \right)  \\
  \pderiv{T_\perp}{t}
  &+ \NL{T_\perp} - B \nabla_\parallel \frac{u_\parallel}{B} + B^2 \nabla_\parallel \frac{q_\perp + u_\parallel}{B^2}  - \left[ \frac{1}{2 L_{ns}} \hat{\nabla}_\perp^2 + \frac{1}{L_{Ts}} \left( 1 + \hat{\hat{\nabla}}_\perp^2 \right) \right] i \omega_\star \Phi \notag \\
  &+ \left( 1 +  \hat{\nabla}_\perp^2 + \hat{\hat{\nabla}}_\perp^2 \right) i \omega_d \Phi + i \omega_d \left( 4 T_\perp + n \right)  + 2 \left| \omega_d \right| \left( \nu_3 T_\parallel + \nu_4 T_\perp \right) = \frac{1}{3} \nu_{ss} \left( T_\parallel - T_\perp \right)  \\
  \pderiv{q_\parallel}{t}
  &+\NL{q_\parallel} + \left( 3 + \beta_\parallel \right) \nabla_\parallel T_\parallel + \sqrt{2} D_\parallel \left| k_\parallel \right| q_\parallel  + i \omega_d \left( -3 q_\parallel - 3 q_\perp + 6 u_\parallel \right)  \notag \\&+ \left| \omega_d \right| \left( \nu_5 u_\parallel + \nu_6 q_\parallel + \nu_7 q_\perp \right) = - \nu_{ss} q_\parallel  \\
  \pderiv{q_\perp}{t}
  &+\NL{q_\perp}+ \nabla_\parallel \left( T_\perp + \frac{1}{2} \hat \nabla_\perp^2 \Phi \right)  + \sqrt{2} D_\perp \left| k_\parallel \right| q_\perp + \left( T_\perp - T_\parallel + \hat{\hat{\nabla}}_\perp^2 \Phi - {1 \over 2} {\hat \nabla}_{\perp}^2 \Phi \right) \nabla_\parallel \ln B  \notag\\
  &+ i \omega_d \left(  -q_\parallel - q_\perp + u_\parallel \right) + \left| \omega_d \right| \left( \nu_8 u_\parallel + \nu_9 q_\parallel + \nu_{10} q_\perp \right) = - \nu_{ss} q_\perp  \label{eq:gf6}
\end{align}
where
\begin{gather*}
  \nabla_\parallel = v_{ts}\uv{b} \cdot \nabla, \qquad b = k_\perp^2 \rho_s^2, \qquad \Phi = \sgam(b) \varphi, \qquad \vPsi = \uv{b} \times\nabla\Phi, \\
  \flr\Phi = b \pderiv{\sgam}{b}\varphi, \qquad \hat{\hat{\nabla}}_\perp^2 \Phi = b \pderiv{^2}{b^2}(b\sgam)\varphi, \\
  i\omega_* =-\nabla \psi \cdot \uv{b} \times \nabla, \qquad i\omega_d = \frac{\tau_s}{Z_s B^2} \uv{b} \times \nabla B \cdot \nabla,
\end{gather*}
%\end{widetext}
and the species thermal velocity $v_{ts}$, gyroradius $\rho_s$, temperature $\tau_s$, charge $Z_s$, equilibrium density and temperature scale lengths $L_{ns}$ and $L_{Ts}$, and collision frequency $\nu_{ss}$ have all been non-dimensionalised. The nonlinear terms are denoted by $\NL{}$ and will be addressed in detail in Appendix~\ref{app:NLPM}; the final form of these terms is given in (\ref{eq:NLn}--\ref{eq:NLqprp}).
The quasineutrality constraint for a single ion species is
\begin{equation}
  n_e = \frac{n}{1+b/2} - \frac{b T_\perp}{2(1+b/2)^2} + (\Gamma_0 - 1) \varphi, \label{qneut} %chktex 3
\end{equation}
where $\Gamma_n(b)= I_n(b)e^{-b}$ and $I_n(b) = i^{-n}J_n(i b)$ is the modified Bessel function.
When electrons are assumed to be adiabatic, which is the case for all results in this paper, we have
\begin{equation}
  n_e = \frac{T_{i}}{T_{e}}(\varphi - \fsa{\varphi}), \label{adiab}
\end{equation}
where $\fsa{\varphi}$ is a flux surface average.

The coefficients $D_\parallel$, $D_\perp$, $\beta_\parallel$, and $\nu_1 -\nu_{10}$ are set by the closures and taken to be the same as in~\citep{Beer1996a}. These closure approximations are carefully chosen to capture important kinetic effects, notably Landau damping, phase mixing from toroidal $\nabla{B}$ and curvature drifts, and finite Larmor radius (FLR) effects.  The resulting gyrofluid model can reproduce the gyrokinetic linear dispersion relation quite accurately.

\section{Closures for nonlinear FLR phase mixing}\label{app:NLPM}
Phase mixing processes, like Landau damping, are fundamentally caused
by the fact that particles have a distribution of velocities. For the
case of Landau damping, the spread in parallel velocities of particles
freely streaming along field lines causes neighboring particles to
move apart. This (linear) phase mixing process smears away spatial
perturbations, even in the collisionless limit, and drives the
formation of small-scales structures in $f(v_\parallel)$, with $\delta
v_\parallel \sim 1/k_\parallel t$.

A similar phase mixing process is associated with the nonlinear term
in the gyrokinetic equation. This term represents random mixing by
gyro-averaged ${\bf E} \times {\bf B}$ flows, and thus produces small
scale structure in space. There is a spread in the gyro-averaged ${\bf
E} \times {\bf B}$ velocities of particles, as higher energy
particles with larger gyroradii average over more fluctuations in the
potential and thereby have a slower ${\bf E} \times {\bf B}$ drift
than lower energy particles with smaller gyroradii. This leads to
phase mixing perpendicular to the field and drives the formation of
structure in $f(v_\perp)$. Thus the nonlinear term simultaneously
produces small scale structures in both physical space and
perpendicular velocity space.

This process was first recognized in the context of gyrofluid closures
by~\cite{Dorland1993a}. Later,~\cite{Schekochihin2009} identified the existence of a
kinetic cascade due to nonlinear phase mixing. They predicted the key
properties of this cascade, and placed it in the context of the
broader concept of entropy cascades. Notably, nonlinear phase mixing
was found to be the dominant method of generating small-scale
structure in velocity space, outpacing Landau damping.
\cite{Tatsuno2012} studied nonlinear phase mixing in the %chktex 2
context of freely decaying turbulence, identifying three regimes of
importance, from collisional to collisionless.
\cite{Howes2011} found numerical evidence supporting %chktex 2
the existence of nonlinear phase mixing and the entropy cascade at
small scales in electromagnetic, kinetic Alfv\'en wave turbulence. The
effects of nonlinear phase mixing have also been observed
experimentally, in laboratory magnetized plasmas
\citep{Kawamori2013} and in the solar wind
\citep{Chen2010}.

We will now derive a gyrofluid closure to model the effects of nonlinear FLR
phase mixing. We start with a simple kinetic problem from which we
derive a damped kinetic response. We then choose a dissipative gyrofluid
closure and fit closure coefficients so that the gyrofluid response closely
matches the kinetic response.

\subsection{Simple kinetic problem}\label{sec:gkadvect}

To illustrate the essence of the nonlinear phase mixing process, we follow%
~\cite{Dorland1993a} and start with a kinetic picture in sheared slab
geometry. We consider a zonal potential that varies sinusoidally in only the
$x$ direction with wavenumber $k_x$: $\varphi = \varphi_{ZF}(x) = {\phi}_{ZF}
\sin (k_x x)$. The zonal $\bf E \times B$ flow is then $\vE = v_{E}(x) \hat{y}
= \pderiv{\varphi_{ZF}}{x} \hat{y} =  k_x \phi_{ZF} \cos(k_x x) \hat{y}$.
Assuming no gradients in the equilibrium $F_0$ and no parallel gradients,
the gyrokinetic equation reduces to a one-dimensional advection equation involving the
nonlinear term,
\begin{align}
  \pderiv{g}{t} + J_0 \left( \frac{k_x v_\perp}{\Omega} \right) v_{E} \pderiv{g}
  {y} = 0. \label{gk_nlpm}
\end{align}
Taking a Maxwellian initial perturbation with a single mode in $y$ with
wavenumber $k_y$, $ g(t=0) = e^{i k_y y} F_M$, the perturbation
then evolves as
\begin{align}
  g(t) &= F_M e^{ik_y [y- J_0(k_x v_\perp/\Omega) v_{E} t]} \simeq
  F_M e^{i k_y (y-v_E t)}e^{i k_y b v_E t v_\perp^2/4 v_t^2},
  \label{eq:F1_nlpm}
\end{align}
where on the right we have expanded to first order in small $b$ by taking
$J_0\simeq1-\frac{b}{4}v_\perp^2$, and
here $b = k_x^2 v_t^2/\Omega^2=k_x^2\rho^2$. In this small-$k_\perp\rho
$ limit we can analytically calculate the kinetic perturbed density response,
given by
\begin{align}
  n_{{kin}}(t) &= \frac{1}{n_0} \int d^3v\  g \simeq  e^{i k_y (y-
  v_{E} t)} \frac{1}{1 - i k_y b v_{E} t/2}\label{nkin}.
\end{align}
Thus we see that the density response decays in time with a long tail that
goes like $1/t$. This is the behavior that we will want to capture with our
fluid closure. Note that we can also numerically integrate the exact kinetic
solution in~\eqref{eq:F1_nlpm} to capture the full $J_0$ effects; for $k_
\perp\rho\leq1$ the exact response is nearly identical to the small-$k_\perp
\rho$ response given in~\eqref{nkin}.

\subsection{Fluid picture}
If we take Laguerre moments in $\mu B=v_\perp^2/2$ of our simple 1D gyrokinetic equation~\eqref{gk_nlpm}, we will see that each Laguerre moment is coupled via the nonlinear term, which is a manifestation of the phase mixing process~\citep{Mandell2018}. Examining the equations for the first two Laguerre moments, $G_{0,0}=n$ and $G_{1,0}=T_\perp$, and again taking the small-$k_\perp\rho$ limit,
we have
\begin{gather}
  \pderiv{n}{t} + \left(1- \frac{b}{2}\right) v_E \pderiv{n}{y} - \frac{b}{2} v_E\pderiv{T_\perp}{y}  = 0, \label{lowb_dens} \\
  \pderiv{T_\perp}{t} + \left( 1- \frac{3b}{2} \right) v_E \pderiv{T_\perp}{y} - \frac{b}{2} v_E\pderiv{n}{y} - \frac{b}{2} v_E\pderiv{G_{2,0}}{y} = 0, \label{lowb}
\end{gather}
where the $G_{2,0}$ Laguerre moment is not evolved but requires closure.
These equations are identical to the Beer equations in the 1D low $k_\perp \rho$ limit, with the exception of the last $G_{2,0}$ term in~\eqref{lowb}. Now that we have identified this extra term, we can generalize these equations to 3D, and to higher $k_\perp \rho$ using the full FLR expressions~\citep{Dorland1993a}:
\begin{gather}
  \pderiv{n}{t} +{\bf v}_\Phi \cdot \nabla n + \left[ {1 \over 2} {\hat \nabla}_{\perp}^2 {\bf v}_{\Phi} \right] \cdot \nabla T_\perp  = 0, \label{genb_dens} \\
  \pderiv{T_\perp}{t} + {\bf v}_\Phi \cdot \nabla T_\perp + \left[ {1 \over 2} {\hat \nabla}_{\perp}^2 {\bf v}_{\Phi} \right] \cdot \nabla n  + \left[ \hat{\hat{\nabla}}_\perp^2 {\bf v}_{\Phi} \right] \cdot \nabla T_\perp +   \left[ {1 \over 2} {\hat \nabla}_{\perp}^2 {\bf v}_{\Phi} \right] \cdot \nabla G_{2,0} = 0. \label{genb_tprp}
\end{gather}

\subsection{Fluid closure}
Now we must find a closure expression for the extra term involving $G_{2,0}$ in~\eqref{genb_tprp}. If we simply set $G_{2,0}=0$, as Beer did, we get an oscillatory (undamped) solution for $n$ and $T_\perp$. In the spirit of the closures pioneered by Hammett and Perkins to model Landau damping~\citep{Hammett1990}, we can instead use dissipative closures to produce fluid density and temperature responses that are damped like the kinetic responses. Thus we will choose a closure of the form
\begin{align}
  G_{2,0}= \frac{\left|[\flr\vPsi] \cdot \nabla\right|}{[\flr\vPsi] \cdot \nabla} \left( \mu_1 n + \mu_2 T_\perp\right),
\end{align}
where we follow~\citep{Beer1996a} and allow each coefficient $\mu$ to have a dissipative and a reactive (non-dissipative) piece, given by
\begin{gather}
  \mu = \mu_r + \mu_i \frac{\left|[\flr\vPsi] \cdot \nabla\right|}{[\flr\vPsi] \cdot \nabla} = (\mu_r,\ \mu_i).
\end{gather}
We set the $\mu_1$ and $\mu_2$ coefficients by numerically minimizing the difference between the kinetic density response~\eqref{nkin} and the fluid density response found by evolving the $n$ and $T_\perp$ equations after inserting the closures. We have found that
$\mu_1=(0.747,-0.078)$ and $\mu_2=(1.368,-2.023)$ produces a fluid response that fits the kinetic response reasonably well for $k_\perp \rho \lesssim 1$.

\subsection{Extension to other moment equations}

One can follow a similar procedure to develop nonlinear phase mixing closures for the remaining gyrofluid equations in the 4+2 model. The $u_\parallel$ and $q_\perp$ equations form a coupled system identical to the system we have studied above, so we can use the same 2-moment closure and simply replace $n$ and $T_\perp$ with $u_\parallel$ and $q_\perp$, respectively. The $T_\parallel$ and $q_\parallel$ equations are also identical; each of these is not coupled to other equations, so the closures that appear in these equations are 1-moment closures. The closure coefficient is found with a similar procedure to the one used above, by fitting the fluid response for $T_\parallel$ to the corresponding kinetic response.

This completes our derivation of a gyrofluid closure to model nonlinear FLR phase mixing. Adding these new terms to the original nonlinear terms, the final gyrofluid nonlinear terms used in \gryfx are given by
\begin{gather}
  \NL{n} = {\bf v}_\Phi \cdot \nabla n + \left[ {1 \over 2} {\hat \nabla}_{\perp}^2 {\bf v}_{\Phi} \right] \cdot \nabla T_\perp \label{eq:NLn}, \\
  \NL{u_\parallel} =  {\bf v}_\Phi \cdot \nabla u_\parallel + \left[ {1 \over 2} {\hat \nabla}_{\perp}^2 {\bf v}_{\Phi} \right] \cdot \nabla q_\perp, \\
  \NL{T_\parallel} = {\bf v}_\Phi \cdot \nabla T_\parallel + \left|\left[\flr\vPsi\right] \cdot \nabla\right|\mu_3 T_\parallel,\\
  \NL{T_\perp} = {\bf v}_\Phi \cdot \nabla T_\perp + \left[ {1 \over 2} {\hat \nabla}_{\perp}^2 {\bf v}_{\Phi} \right] \cdot \nabla n + \left[ \hat{\hat{\nabla}}_\perp^2 {\bf v}_{\Phi} \right] \cdot \nabla T_\perp + \left|\left[\flr\vPsi \right]\cdot \nabla\right| \left(\mu_1 n + \mu_2 T_\perp \right),  \\
  \NL{q_\parallel} = {\bf v}_\Phi \cdot \nabla q_\parallel + \left|\left[\flr\vPsi\right] \cdot \nabla\right|\mu_3 q_\parallel, \\
  \NL{q_\perp} = {\bf v}_\Phi \cdot \nabla q_\perp + \left[ {1 \over 2} {\hat \nabla}_{\perp}^2 {\bf v}_{\Phi} \right] \cdot \nabla u_\parallel + \left[ \hat{\hat{\nabla}}_\perp^2 {\bf v}_{\Phi} \right] \cdot \nabla q_\perp + \left|\left[\flr\vPsi\right] \cdot \nabla\right| \left(\mu_1 u_\parallel + \mu_2 q_\perp\right) \label{eq:NLqprp},
\end{gather}
where $\NL{m}$ represents the nonlinear terms in the $m$ moment equation, and the absolute value terms comprise our new closure terms with closure coefficients $\mu_1=(0.747,-0.078),\ \mu_2=(1.368,-2.023)$, and $\mu_3=(0.456,-0.724)$. This set of closures is similar to those presented in~\citep{Dorland1993a}, but that model required higher order terms of order $k_\perp^4\rho^4$ in the $n$ and $u_\parallel$ equations, which is beyond the order of accuracy of the usual FLR terms. We avoid this by including closure terms proportional to $n$ and $u_\parallel$ in the $T_\perp$ and $q_\perp$ equations, respectively.

In the derivation of these new closures, we only considered nonlinear FLR phase mixing from a static, 1D potential. In an evolving 3D system, the phase mixing process is much more complicated, and our rough model may overestimate or underestimate the amount of phase mixing. Nonetheless, our model will capture the correct scaling of the mixing, producing physically motivated damping at large $\varphi$ and high $k_\perp$. In this way our new terms can be interpreted as a type of hyperviscosity, but one that damps at the $\rho$ scale as opposed to conventional hyperviscosity models that damp at the grid scale.

Finally we must address how we evaluate and implement a term of the form $\mathcal{P}=|{\bf v}\cdot \nabla|M$. Since we use a Fourier spectral representation for the equations, we are interested in the Fourier transform of $\mathcal{P}$. Denoting the Fourier transform of a quantity $x$ as $\hat{x}_k$, and defining $\NL{}={\bf v}\cdot\nabla M$, we evaluate the closure term as
\begin{gather}
  \hat{\mathcal{P}}_k =\widehat{(|{\bf v}\cdot \nabla|M)}_k= | \hat{\mathcal{N}}_k | \frac{\hat{M}_k}{|\hat{M}_k|},
\end{gather}
where all operations are performed in Fourier space, and $\hat{\mathcal{N}}_k$ can be calculated pseudospectrally in the same manner as the usual nonlinear terms.

\section{Hybrid gyrokinetic-gyrofluid zonal flow model}\label{app:zonal}
A major drawback of the Beer gyrofluid model is the inability to accurately model zonal flows.
These nonlinearly-driven sheared poloidal {\bf{E}} $\times$ {\bf{B}} flows have been shown to play a key role in determining the turbulence saturation level. 
Therefore inaccuracy in zonal flow dynamics has been one of the main sources of disagreement between gyrofluid and gyrokinetic turbulence models.\footnote{This is not to say that zonal flow dynamics is the only source of disagreement. 
Our results indicate that even with accurate zonal flow dynamics, a model of nonlinear phase mixing is required to produce the agreement between the gyrofluid and gyrokinetic models shown in Figure~\ref{fig:gryfxgs2comp}.} Attempts were made, with limited success, to modify the gyrofluid closures~\citep{Beer1998} to capture the linearly undamped component of the flows derived by~\cite{Rosenbluth1998}. 
These closure modifications had limited success~\citep{Dimits2000}, and also relied on simplifying assumptions about the magnetic geometry, which would not be able to capture the dependence of zonal flow dynamics on shaping~\citep{Xiao2007}. Further, accurately modeling zonal flows involves resolving sharp features in the distribution function in the $v_\parallel$ direction resulting from trapped particle dynamics, which in a moment approach would require much higher Hermite moments in $v_\parallel$ than used in the Beer model~\citep{Mandell2018}.

Instead of seeking more complicated gyrofluid closure modifications to improve zonal flow accuracy, we avoid this issue by employing a hybrid approach in \gryfx: we evolve the zonal flow modes with a fully gyrokinetic model (using the gyrokinetic code \gstwo), while continuing to evolve the non-zonal modes with the gyrofluid model given by equations (\ref{eq:gf1}--\ref{eq:gf6}) above. Because we use a Fourier spectral representation for the perpendicular configuration space discretization in \gryfx, and because the Fourier modes only interact via the nonlinearity,
we can easily choose an alternative algorithm for the linear evolution of the zonal ($k_y=0$) modes. In order to nonlinearly couple the zonal and non-zonal modes, we must first be able to transform the gyrokinetic distribution function into gyrofluid moments, and vice versa. The transformation from the gyrokinetic distribution function to gyrofluid moments is simply the process of taking velocity moments, which can also be interpreted as projecting the distribution function onto a Laguerre-Hermite basis as shown in equation~\eqref{eq:project}. For the inverse transformation, from gyrofluid moments to the gyrokinetic distribution function, we can expand the distribution function in the Laguerre-Hermite basis as~\citep{Mandell2018}
\begin{align}
  \frac{g}{F_0} &= \sum_{\ell=0} \sum_{m=0}  \frac{(-1)^\ell}{\sqrt{m!}}\ \mathrm{L}_{\ell}(\mu B)\  \mathrm{He}_m(v_\parallel)\ G_{\ell,m}\\%chktex 3
  &\approx n + v_\parallel u_\parallel + \frac{1}{2}\left(v_\parallel^2-1\right)T_\parallel + \left(\mu B-1\right)T_\perp \notag \\&\qquad+ \frac{1}{2}\left(\frac{v_\parallel^3}{3}-v_\parallel\right)q_\parallel + v_\parallel\left(\mu B - 1\right)q_\perp
\end{align}
where in the second line we have explicitly expressed the expansion in terms of the Beer gyrofluid moments and truncated.

Following the same logic, we can expand the gyrokinetic nonlinear term in terms of the gyrofluid nonlinear terms given in (\ref{eq:NLn}--\ref{eq:NLqprp}):
\begin{align}
  \NL{gk} &= \uv{b}\times \nabla \gyroR{\varphi} \cdot \nabla h =  \uv{b}\times \nabla \gyroR{\varphi} \cdot \nabla g \notag \\
  &\approx  \NL{n} + v_\parallel \NL{u_\parallel} + \frac{1}{2}\left(v_\parallel^2-1\right)\NL{T_\parallel} + \left(\mu B-1\right)\NL{T_\perp} \notag \\&\qquad+ \frac{1}{2}\left(\frac{v_\parallel^3}{3}-v_\parallel\right)\NL{q_\parallel} + v_\parallel\left(\mu B - 1\right)\NL{q_\perp}, \label{eq:gknl}
\end{align}
Thus we can construct the gyrokinetic nonlinear term for the $k_y=0$ zonal modes from the $k_y=0$ component of the six gyrofluid nonlinear terms. The full hybrid algorithm then proceeds as follows.\\
\begin{enumerate}
  \item Calculate moments of the $k_y=0$ component of the gyrokinetic distribution function (\emph{e.g.} via~\eqref{eq:project}).
  \item Evaluate the six gyrofluid nonlinear terms, (\ref{eq:NLn}--\ref{eq:NLqprp}), for all Fourier modes.
  \item Evaluate the $k_y=0$ component of the gyrokinetic nonlinear term via~\eqref{eq:gknl}.\label{gknl}
  \item Evolve the $k_y\neq0$ modes with equations (\ref{eq:gf1}--\ref{eq:gf6}).\label{gfevolve}
  \item Evolve the $k_y=0$ modes with equation~\eqref{eqn:gk} using the nonlinear term from (\ref{gknl}).\\\label{gkevolve}
\end{enumerate}
As mentioned above, we couple to the gyrokinetic code \gstwo to evolve the gyrokinetic equation in step (\ref{gkevolve}). Further, we do not include the nonlinear phase mixing closure terms in the $k_y=0$ component of the gyrofluid nonlinear terms. This is in part due to the fact that introducing damping from closure models can suppress the zonal flow residual (which is precisely the reason we moved to gyrokinetic evolution of the zonal modes). Thus in our model, the nonlinear drive for the zonal flows comes only from the lowest several moments of the distribution function; this approximation is justified by the results of~\citep{Rogers2000}.

Finally, note that steps (\ref{gfevolve}) and (\ref{gkevolve}) can be executed in parallel. In \gryfx we take advantage of GPU-CPU concurrency by simultaneously evolving the non-zonal gyrofluid equations on the GPU and the zonal gyrokinetic equation on the CPU.\@
Using a single GPU and $\sim16$ CPU cores (a common supercomputer node configuration), the two steps take roughly the same amount of wall clock time. This means that in our scheme the additional cost (in terms of wall clock time) of using a fully gyrokinetic model for the zonal flows is minimal.

\section{Practical Considerations}\label{app:practicalconsiderations}

\begin{figure}
  \begin{center}
    \includegraphics[width=0.98\textwidth]{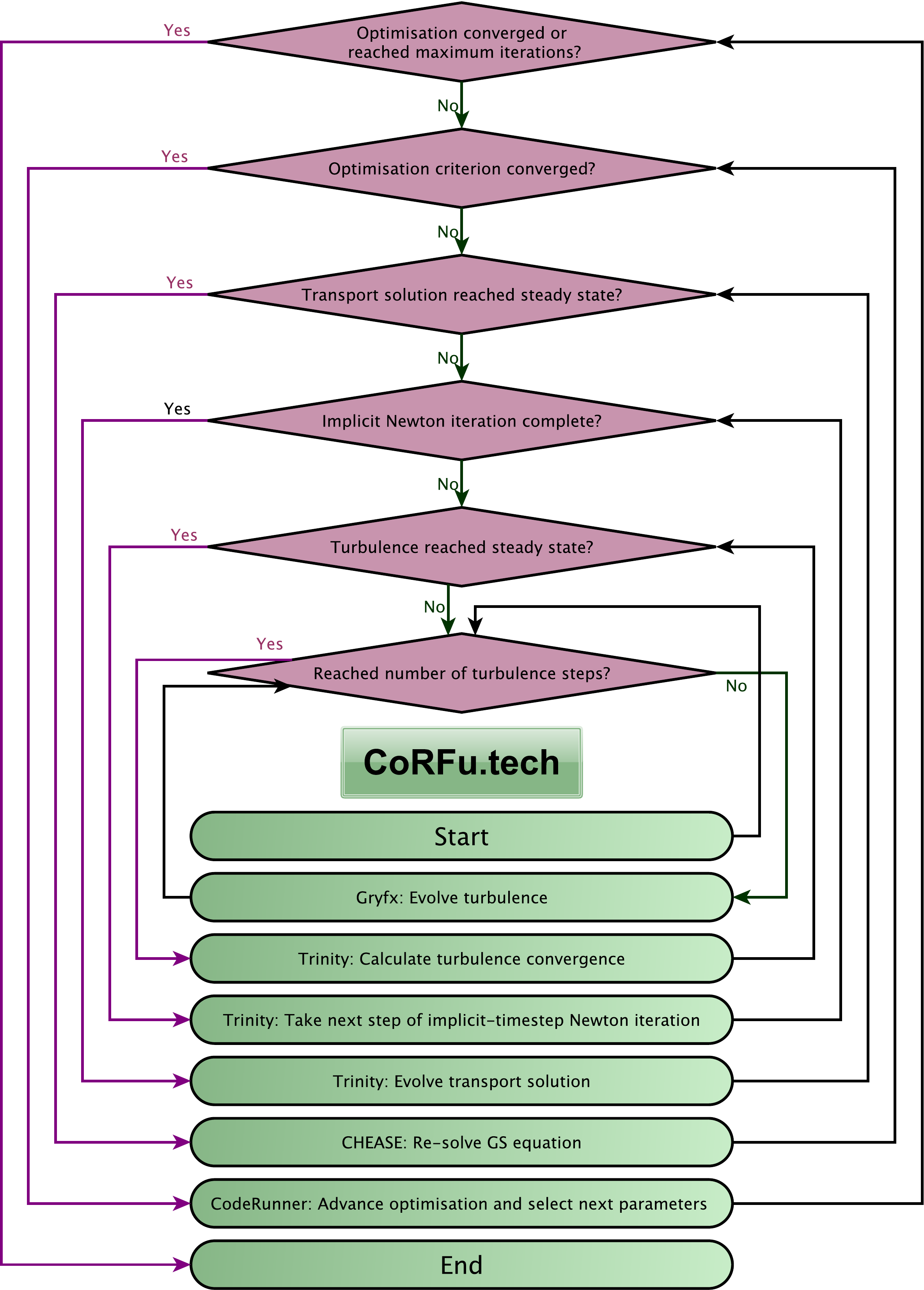}
  \end{center}
  \caption{Schematic of \corfu operation showing the hierarchy of decision loops which must be walked through.}\label{fig:corfu_loops}
\end{figure}

The presentation of equations and the methodology given above has,
for reasons of clarity and brevity, skipped over some of the most thorny
issues encountered during the current study.
However, we believe that a discussion of them should be present
in this work, and we hope that by including it here as an appendix, 
we can save at least
some time and effort for future researchers.
In this discussion it helps to think of the operation of \corfu as a series of six nested loops, as illustrated in \figref{corfu_loops}. Within each of these loops issues can arise.
The principle practical challenges encountered in the construction and use of the 
%CoTrOp CTO CoTO CoRuTriO CoRnTrio CoRnTri CRuTOp CRuTiO CoReTriO CoRuTO CoRuTOp CRunT CoRupT CoFOT CRuFraT CoRFu TriP CoReTFvO  
\corfu framework
(independent of the challenges in constructing its individual components such as \trinity or GRYFX)
were:

\begin{enumerate}
  \item incorporating evolution of the magnetic equilibrium,
  \item determining whether a particular turbulence calculation had reached steady state,
  \item encountering failures of either the turbulence code or the Grad-Shafranov (GS) code,
  \item dealing with turbulent thresholds,
  \item making efficient use of resources, and
  \item choosing resolution parameters for the turbulence code.
  %\item software choices and engineering, robustness in the face of hardware failure or software bugs.
\end{enumerate}

In the following sections we discuss each of these challenges.

\subsection{Incorporating evolution of the magnetic equilibrium}

The \trinity transport solver has the ability to evolve the magnetic equilibrium internally. 
However, at present it is not able to treat the evolution of the magnetic equilibrium implicitly,
as it does the profiles of density, temperature and flow.
Therefore it would be necessary to use a much smaller transport step to avoid numerical instability.

The advantage of this study is that we are only seeking steady-state transport and magnetic
equilibrium solutions.
Thus, provided \trinity discovers a solution in which both the profiles and the equilibrium do 
not vary in time, this solution will be self-consistent and thus acceptable.
This allows us to separate the evolution of the Grad-Shafranov solution from \trinity, 
as is illustrated in \figref{corfu_loops}.
In effect, \trinity evolves using a fixed magnetic equilibrium until the profiles stop changing.
At this point the new pressure profile is passed to \chease which recalculates the equilibrium.
\trinity is then re-run with new equilibrium and the cycle is repeated until the optimisation
criterion converges to within a specified tolerance (see \appref{turbulenceconvergence}).

\subsection{Determining whether a particular turbulence calculation had reached steady state}%
\label{app:turbulenceconvergence}

This is an extensive subject fraught with complication. Not only is it hard to automate 
what enters into human judgement in these cases (in many cases the best way of determining
whether a simulation has reached saturation is to ``eyeball'' the time traces)
but there are cases when even the best of classification systems would fail 
(e.g.\ the case of slow zonal flow amplitude rise in ETG turbulence;~\cite{Colyer2017}).

The difficulty proceeds from having no \emph{a priori} knowledge of the timescales
of a given situation: in particular, those of the linear growth, time to
saturation, and the longest 
timescales in the saturated state (usually related to zonal flows).
A typical dilemma might be: is the heat flux no longer changing because

\begin{enumerate}
  \item the simulation has reached a saturated nonlinear state, or
  \item the system was stable and all modes died away to a constant noise level, or
  \item the growth rate was so small that the change in amplitude was undetectable on
    the timescale allotted?
\end{enumerate}

The authors of this paper have experimented with many techniques, including
rolling time averages, moving averages and frequency analysis. 
In this work,  a simpler approach was used
which represented a balance between the competing objectives of certainty,
efficiency, and robustness.

\begin{figure}
  \begin{center}
    \includegraphics[width=0.98\textwidth]{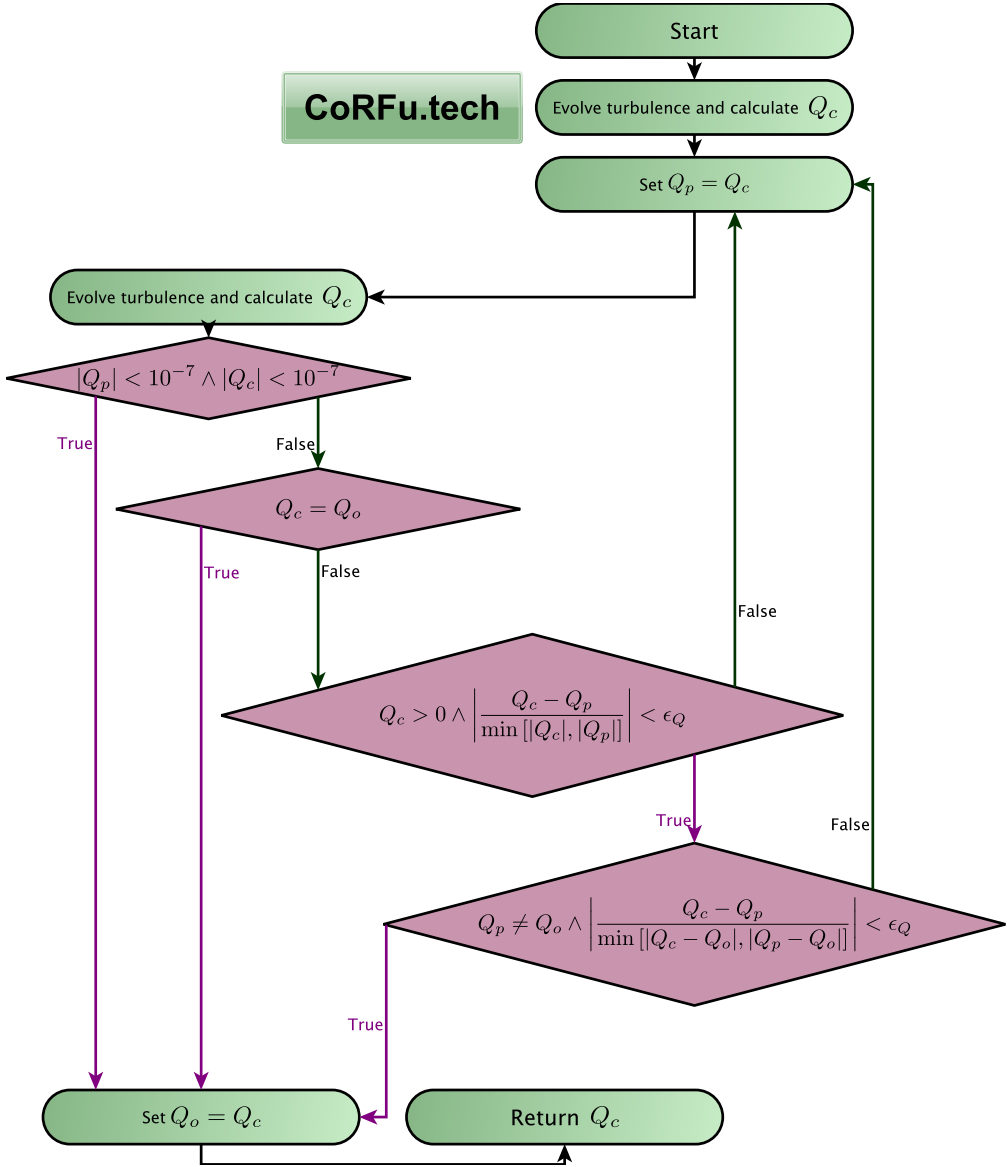}
  \end{center}
  \caption{Schematic of the \corfu turbulence code driver showing the logic used to determine if a turbulence calculation is saturated ($Q_o$ is the heat flux from the previous iteration, $Q_p$ is the heat flux from the previous stage of the turbulence calculation, and $Q_c$ is the heat flux from the current stage).}\label{fig:saturation}
\end{figure}

Each turbulence calculation (that is, for each radial location,
each element of the Jacobian and each transport iteration)
was broken up into a number of stages, with
all fields cached between stages so that the calculation could 
resume instantly from the same state.
The number of time steps in each stage was chosen in combination with the timestep
so that each stage lasted $O\left( 10 \right)$ turnover times ($a/v_{thi}$).
The turbulence code driver then ran the calculation for two initial 
stages, and then proceeded to run each calculation according to the decision
tree presented in \figref{saturation}.
This decision tree is extremely restrictive about allowing a simulation
to be declared converged. 
This choice was motivated by the observation that
a bad value for the heat flux incurred such a great time penalty---%
by disrupting the sensitive implicit Newton iteration---%
that running \gryfx for a little longer than was typically 
necessary saved time in the long run.

\subsection{Encountering failures of either Newton iteration, the turbulence code or the GS~solver}\label{app:failures}

It is of course possible for any component of \corfu to experience a failure during the very long operation typically required.
This could be due to hardware failures, file corruptions, and so on.
However, the three most common failures that were encountered were 
\begin{itemize}
  \item failure of the Newton iteration to converge due to overlarge timestep
    or a bad flux value,
  \item numerical instability from an overly large timestep in \gryfx, 
    and 
  \item failure of \chease to converge on a solution, usually as a result of a
    non-monotonic pressure profile but sometimes from straying to an extreme location
    in parameter space. 
\end{itemize}

In the case of a failure of the Newton iteration, one likely cause was 
the inherent noise in the turbulent fluxes. 
In particular, this was a problem if a large change in a gradient 
(e.g.\ the temperature gradient $\kappa_T=R/L_T$)
produced only a small change in a transport coefficient 
(e.g.\ the heat flux $Q$).
Since we are inverting a Jacobian which essentially depends on quantities
such as $\Delta Q/ \Delta \kappa_T$, if the true value of $\Delta Q$ is small,
and there is noise, of the same magnitude, which contrives to give $\Delta Q$
close to zero, there is an artificial stationary point created in the search 
domain which will lead the Newton iteration to take a large jump incorrectly.
In the future, it is planned to implement a Levenberg-Marquardt mixed search
which will  be more resistant to this; for the current study it was necessary 
to be very stringent with the turbulence convergence conditions, and to use
a large, fixed value of $\Delta \kappa_T = 0.6$. The reasoning behind this was
that since $Q\propto\kappa_{T}^{3}$ \citep{Barnes2011a}, a constant value of
$\Delta \kappa_T$ would continue to produce a respectable $\Delta Q$ as $Q$ tended
to 0, but would not produce an excessively large $\Delta Q$ far from the threshold.

With this implemented, a remaining likely cause of a failure
was %likely to e
a real stationary point in the search domain, and the solution was to make the transport
timestep smaller. In the next section we will discuss this in more detail with
regard to efficiency. However, there was another consideration when resetting the timestep:
what to do about the state of the turbulence calculation. 

If the Newton iteration had deviated very far from the previous profiles it might 
be reasonable to assume that since the turbulent state at the end of the failed iteration
was very far from that at the beginning, it would be better to start the turbulence
calculation from noise to avoid the turbulence displaying violent transients
(particularly if the end of the failed iteration was in a strongly-driven regime).
In fact, since (because of stiff transport) the steady-state profiles were
never very far from marginal (with low growth rates), such a strategy would incur
a large time penalty and run the risk of mistaking a low-growing turbulent case for
a stable case.

In the case of \gryfx, a timestep was chosen that was sufficiently small for almost all
cases. The turbulence calculation driver would typically start a \gryfx simulation stage
using the state of the previous simulation stage (since either it was continuing
a calculation till convergence, or it was starting a new calculation with similar
parameters to the previous); however this behaviour was changed if necessary
(note that resetting the initial condition close to threshold when the growth rate 
was low incurred a significant time penalty).
Otherwise, the following steps were taken to mitigate problems:

\begin{itemize}
  \item If the previous \trinity Newton iteration did not converge, values of fields and moments
    were kept but time averages of fluxes were reset.
  \item If the previous fluxes were very small ($<10^{-7}$), all values were reset.
  \item If the previous fluxes were either very large ($>1e10$) or had \texttt{NaN}
    values, all values were reset. Typically this was because a quirk of the initial
    condition pushed the turbulence amplitudes to values too large for the timestep and could
    be fixed by resetting, but 
    if this occurred repeatedly the optimisation would be halted and require intervention.
  \item If fewer than 3 turbulence stages had been run the moving averages would be reset
    to eliminate initial transients.
  \item If greater than 3 turbulence stages had been run, the timescale of the moving
    average would gradually increase in successive stages (proportional to the total
    turbulence time elapsed in these successive stages) to force eventual convergence 
    of the heat flux. This was a necessary to ensure that the transport calculation
    never became stuck, and had no impact on the results, once again because only
    steady-state solutions were required.
\end{itemize}

In the case of the Grad-Shafranov solver \chease, since failures were often caused
by an unconverged pressure profile, \corfu first attempted to run \trinity for
an extra period of time
using the previous equilibrium. 
Occasionally in preliminary studies (though not in the study reported here)
the GS solver failed because the control parameters strayed into a region
of parameter space where a solution could not be found. 
In such cases a highly unfavourable value was returned to prevent
the algorithm pursuing that search direction.

\subsection{Dealing with turbulent thresholds}

To a user of the \corfu framework, it is an inconvenient reality that
optimal solutions typically lie close to the zero-turbulence manifold. % (because,
%roughly speaking, the higher the threshold, the stiffer the transport).
This means that the transport solver \trinity will spend most of its operation 
hovering around the turbulent threshold. 
This is a problem because of the extra time required to calculate the 
turbulent transport.
If the previous transport iteration was below the threshold and 
the subsequent iteration is likewise, the turbulence calculation
will end rapidly because any initial noise will die away. 
If the previous iteration was above threshold,
and the subsequent one is too, then the turbulence will rapidly
adjust to the new drive parameters.
However, if the previous iteration was below threshold and the subsequent one
above, (i.e\ the transport solver crosses the zero-turbulence
manifold from below) the turbulence calculation will not converge
until amplitudes have grown linearly to saturation amplitudes.
Since we are close to threshold with small growth rates, this
may take a very long time. 
Equally unpleasant is when the zero-turbulence manifold is crossed
from above: in this case the free energy of the previous vigorous
turbulence must be absorbed by only weakly damped modes.

The strategy for dealing with the zero-turbulence manifold 
was to adjust the way that \trinity calculated the derivatives
of fluxes (e.g.\ $Q$) with respect to profile gradients 
(e.g.\  $\kappa_T$).
Effectively we needed to make sure that, as far as possible,
the lower part of the derivative stencil stays below the 
threshold, and the higher part stays above the threshold. 
This is achieved firstly by ensuring that the stencil is always
fairly broad (at the cost of reducing the convergence
rate of the Newton iteration and hence the size of the
timestep).
Fortunately, this requirement coincided with the
requirement described in \appref{failures} for a large
$\Delta Q$.
The second way to achieve this is to set a tight 
convergence criterion for the Newton iteration to make sure that
it rarely artificially (that is, at odds with the continuous time-evolution)
crossed the threshold (in 
terms of \trinity parameters, \texttt{errflr} was set to 0.02
and \texttt{flrfac} was set to 4.0).

%After experimenting with absolute and proportional 
%values it was determined 

\subsection{Making efficient use of resources}

Although \gryfx is a very fast and efficient code, it is very important,
given the number of turbulence calculations carried out, to strictly
control the cost of the optimisation study.
For many reasons, as given above, it is necessary to be overly 
cautious and run the turbulence simulations until a 
strict set of criteria have been satisfied. 
Nonetheless, it will always be the case that during every iteration
of the transport calculation some flux calculations will converge
before others. Therefore the flux driver is set up so that the optimisation
can run on an arbitrary number of GPUs, and that if the number of 
GPUs is less than the number 
of flux calculations carried out at a time, then 
at each stage of the flux calculation the GPUs will 
cycle through each of the flux calculations requested by 
the driver. As the number of stages increases, more and
more of the turbulence calculations will have converged and so 
the GPUs will cycle more quickly through the remaining ones. 
In the current study, with eight flux calculations per transport iteration, 
running on four GPUs, this arrangement reduced the simulation cost by
close to 50\%.

An equally important consideration with regard to cost is the choices
made for iteration and time stepping parameters within \trinity.
When seeking a steady state solution, a certain amount of time 
will need to elapse while the profiles adjust to the new parameters
provided by the optimisation driver.
A first assumption might be that increasing the size of the timestep
will reduce the cost of evolving the profiles for that time (remembering
that the \trinity algorithm is implicit and therefore stable for
large timesteps).
However, the larger the timestep, the greater the difference between
the profiles at the end and the profiles at the beginning of the timestep,
and thus the greater the chance that the Newton search will not be
able to land upon the future profiles. 
If this occurs, the timestep must be reduced and the step repeated,
effectively doubling the cost of that timestep. 
Increasing the maximum number of steps in the Newton iteration
would increase the chances of it reaching a solution; 
however, it would also increase
the cost should the iteration still fail. 

By trial and error, the following choices were found
to enable efficient and robust operation.
\begin{itemize}
  \item The threshold for rejecting a Newton iteration was set 
    high (\texttt{errtol}=0.1) to avoid the cost of repeating 
    an iteration (this value was still low enough to avoid
    wacky iterations triggered by a stationary point).
  \item The threshold for saying that the Newton iteration
    was converged and therefore fewer than the allowed number
    of steps were needed was set low (\texttt{errflr}=0.02)
    so that the iteration was always as accurate as possible,
    leading to an accurate time evolution.
   \item The threshold for increasing the timestep was set
     very low (\texttt{flrfac}=4 so that the threshold was
     0.02/4) so that the timestep was only increased if the
     iterations were proceeding very smoothly.
   \item The factor by which the timestep was reduced after
     a failed iteration was set high (\texttt{deltadj}=4)
     to avoid repeated failed iterations. For example, when crossing a 
     turbulent threshold it was often necessary to rapidly
     reduce the transport timestep: if \texttt{deltadj} was
     small, it would take a long time before the timestep
     was reduced sufficiently.
   \item The maximum allowed number of steps in the Newton iteration
     was set to 4. This represented a balance between
     giving the search the greatest chance of succeeding and reducing
     the cost if the search eventually failed.
\end{itemize}

\subsection{Choosing resolution parameters for the turbulence code}

The resolution required for the turbulence calculation, including
perpendicular box size, perpendicular and parallel resolutions,
and timestep, vary significantly with the values of the driving
parameters and the magnetic geometry.

In the present case, the GPUs that were available
for the calculation were sufficiently powerful to allow 
the increase of the \gryfx resolution to a size that was
more than adequate for the majority of simulations
(as determined by manual inspection of a sample of turbulence
simulations), 
without an unmanageable increase in cost. 

There are \textit{a posteriori} ways of checking that resolutions
were sufficient, such as examining turbulent spectra,
which would allow one to avoid such a blanket cost increase.
It is also, of course, possible to check that the output quantities
such as heat flux do not change with increased resolution.
However, the automation of such investigations, which are usually
carried out by hand for individual turbulence calculations,
has not yet been implemented in the \corfu framework, and would
constitute significant additional research.

%(We note that
 %a GPU is a discrete computational unit which can only run
 %exactly one \gryfx simulation at a time; thus a smaller
 %simulation may lea

%\subsection{Software choices and engineering, robustness in the face of hardware failure or software bugs}

%\clearpage
\bibliographystyle{jpp}
\bibliography{library,library_noah}

\begin{thebibliography}{75}
\expandafter\ifx\csname natexlab\endcsname\relax\def\natexlab#1{#1}\fi
\def\au#1{#1} \def\ed#1{#1} \def\yr#1{#1}\def\at#1{#1}\def\jt#1{\textit{#1}}
  \def\bt#1{#1}\def\bvol#1{\textbf{#1}} \def\vol#1{#1} \def\pg#1{#1}
  \def\publ#1{#1}\def\arxiv#1{#1}\def\org#1{#1}\def\st#1{\textit{#1}}

\bibitem[Abel \& Cowley(2013)]{Abel2013a}
{\sc \au{Abel, I.~G.} \& \au{Cowley, S.~C.}} \yr{2013}  \at{{Multiscale
  gyrokinetics for rotating tokamak plasmas: II. Reduced models for electron
  dynamics}}.  \jt{New Journal of Physics}  \bvol{15},  \arxiv{arXiv:
  1210.1417}.

\bibitem[Abel {\em et~al.\/}(2013)Abel, Plunk, Wang, Barnes, Cowley, Dorland \&
  Schekochihin]{Abel2013}
{\sc \au{Abel, I.~G.}, \au{Plunk, G.~G.}, \au{Wang, E.}, \au{Barnes, M.},
  \au{Cowley, S.~C.}, \au{Dorland, W.} \& \au{Schekochihin, A.~A.}} \yr{2013}
  \at{{Multiscale gyrokinetics for rotating tokamak plasmas: fluctuations,
  transport and energy flows.}}  \jt{Reports on progress in physics. Physical
  Society (Great Britain)}  \bvol{76}~(11),  \pg{116201}.

\bibitem[Abramson {\em et~al.\/}(2006)Abramson, Peachey \& Lewis]{Abramson2006}
{\sc \au{Abramson, D.}, \au{Peachey, T.} \& \au{Lewis, A.}} \yr{2006}
  \at{{Model Optimization and Parameter Estimation with Nimrod / O}}.
  \jt{Proceedings of the 6th international conference on Computational Science
  (ICCS'06)}  \bvol{1},  \pg{720--727}.

\bibitem[Barnes {\em et~al.\/}(2010)Barnes, Abel, Dorland, Görler, Hammett \&
  Jenko]{Barnes2010}
{\sc \au{Barnes, M.}, \au{Abel, I.~G.}, \au{Dorland, W.}, \au{Görler, T.},
  \au{Hammett, G.~W.} \& \au{Jenko, F.}} \yr{2010}  \at{{Direct multiscale
  coupling of a transport code to gyrokinetic turbulence codes}}.  \jt{Physics
  of Plasmas}  \bvol{17}~(5),  \pg{056109},  \arxiv{arXiv: 0901.2868}.

\bibitem[Barnes {\em et~al.\/}(2011{\natexlab{{\em a\/}}})Barnes, Parra,
  Highcock, Schekochihin, Cowley \& Roach]{Barnes2011}
{\sc \au{Barnes, M.}, \au{Parra, F.~I.}, \au{Highcock, E.~G.},
  \au{Schekochihin, A.}, \au{Cowley, S.~C.} \& \au{Roach, C.~M.}}
  \yr{2011{\natexlab{{\em a\/}}}}  \at{{Turbulent Transport in Tokamak Plasmas
  with Rotational Shear}}.  \jt{Physical Review Letters}  \bvol{106}~(17),
  \pg{1--4}.

\bibitem[Barnes {\em et~al.\/}(2011{\natexlab{{\em b\/}}})Barnes, Parra \&
  Schekochihin]{Barnes2011a}
{\sc \au{Barnes, M.}, \au{Parra, F.~I.} \& \au{Schekochihin, a.~a.}}
  \yr{2011{\natexlab{{\em b\/}}}}  \at{{Critically Balanced Ion Temperature
  Gradient Turbulence in Fusion Plasmas}}.  \jt{Physical Review Letters}
  \bvol{107}~(11),  \pg{115003}.

\bibitem[Beer \& Hammett(1998)]{Beer1998}
{\sc \au{Beer, M.} \& \au{Hammett, G.}} \yr{1998}  \at{The dynamics of
  small-scale turbulence-driven flows}.  \jt{Varenna Proceedings}  \pg{pp.
  1--11}.

\bibitem[Beer {\em et~al.\/}(1995)Beer, Cowley \& Hammett]{Beer1995}
{\sc \au{Beer, M.~A.}, \au{Cowley, S.~C.} \& \au{Hammett, G.~W.}} \yr{1995}
  \at{{Field‐aligned coordinates for nonlinear simulations of tokamak
  turbulence}}.  \jt{Physics of Plasmas}  \bvol{2}~(7),  \pg{2687--2700}.

\bibitem[Beer \& Hammett(1996)]{Beer1996a}
{\sc \au{Beer, M.~A.} \& \au{Hammett, G.~W.}} \yr{1996}  \at{{Toroidal
  gyrofluid equations for simulations of tokamak turbulence}}.  \jt{Physics of
  Plasmas}  \bvol{3}~(11),  \pg{4046}.

\bibitem[Bourdelle {\em et~al.\/}(2016)Bourdelle, Citrin, Baiocchi, Casati,
  Cottier, Garbet \& Imbeaux]{Bourdelle2016}
{\sc \au{Bourdelle, C.}, \au{Citrin, J.}, \au{Baiocchi, B.}, \au{Casati, A.},
  \au{Cottier, P.}, \au{Garbet, X.} \& \au{Imbeaux, F.}} \yr{2016}  \at{{Core
  turbulent transport in tokamak plasmas: bridging theory and experiment with
  QuaLiKiz}}.  \jt{Plasma Physics and Controlled Fusion}  \bvol{58}~(1),
  \pg{014036}.

\bibitem[Budny(2009)]{Budny2009}
{\sc \au{Budny, R.~V.}} \yr{2009}  \at{{Comparisons of predicted plasma
  performance in ITER H-mode plasmas with various mixes of external heating}}.
  \jt{Nuclear Fusion}  \bvol{49}~(8),  \pg{85008}.

\bibitem[Burrell(1997)]{Burrell1997}
{\sc \au{Burrell, K.~H.}} \yr{1997}  \at{{Effects of ExB velocity shear and
  magnetic shear on turbulence and transport in magnetic confinement devices}}.
   \jt{Physics of Plasmas}  \bvol{4}~(5),  \pg{1499--1518}.

\bibitem[Chang \& Hinton(1982)]{Chang1982}
{\sc \au{Chang, C.~S.} \& \au{Hinton, F.~L.}} \yr{1982}  \at{{Effect of
  impurity particles on the finite-aspect ratio neoclassical ion thermal
  conductivity in a tokamak}}.  \jt{Physics of Fluids}  \bvol{25}~(1493),
  \pg{3314}.

\bibitem[Chen {\em et~al.\/}(2010)Chen, Wicks, Horbury \&
  Schekochihin]{Chen2010}
{\sc \au{Chen, C.}, \au{Wicks, R.}, \au{Horbury, T.} \& \au{Schekochihin, A.}}
  \yr{2010}  \at{Interpreting power anisotropy measurements in plasma
  turbulence}.  \jt{The Astrophysical Journal Letters}  \bvol{711}~(2),
  \pg{L79}.

\bibitem[Citrin {\em et~al.\/}(2015{\natexlab{{\em a\/}}})Citrin, Breton,
  Felici, Imbeaux, Aniel, Artaud, Baiocchi, Bourdelle, Camenen \&
  Garcia]{Citrin2015a}
{\sc \au{Citrin, J.}, \au{Breton, S.}, \au{Felici, F.}, \au{Imbeaux, F.},
  \au{Aniel, T.}, \au{Artaud, J.~F.}, \au{Baiocchi, B.}, \au{Bourdelle, C.},
  \au{Camenen, Y.} \& \au{Garcia, J.}} \yr{2015{\natexlab{{\em a\/}}}}
  \at{{Real-time capable first principle based modelling of tokamak turbulent
  transport}}.  \jt{Nuclear Fusion}  \bvol{55}~(9),  \pg{92001},  \arxiv{arXiv:
  1502.07402v1}.

\bibitem[Citrin {\em et~al.\/}(2015{\natexlab{{\em b\/}}})Citrin, Garcia,
  G{\"{o}}rler, Jenko, Mantica, Told, Bourdelle, Hatch, Hogeweij, Johnson,
  Pueschel \& Schneider]{Citrin2015}
{\sc \au{Citrin, J.}, \au{Garcia, J.}, \au{G{\"{o}}rler, T.}, \au{Jenko, F.},
  \au{Mantica, P.}, \au{Told, D.}, \au{Bourdelle, C.}, \au{Hatch, D.~R.},
  \au{Hogeweij, G. M.~D.}, \au{Johnson, T.}, \au{Pueschel, M.~J.} \&
  \au{Schneider, M.}} \yr{2015{\natexlab{{\em b\/}}}}  \at{{Electromagnetic
  stabilization of tokamak microturbulence in a high- $\beta$ regime}}.
  \jt{Plasma Physics and Controlled Fusion}  \bvol{57}~(1),  \pg{014032}.

\bibitem[Citrin {\em et~al.\/}(2014)Citrin, Jenko, Mantica, Told, Bourdelle,
  Dumont, Garcia, Haverkort, Hogeweij, Johnson \& Pueschel]{Citrin2014}
{\sc \au{Citrin, J.}, \au{Jenko, F.}, \au{Mantica, P.}, \au{Told, D.},
  \au{Bourdelle, C.}, \au{Dumont, R.}, \au{Garcia, J.}, \au{Haverkort, J.},
  \au{Hogeweij, G.}, \au{Johnson, T.} \& \au{Pueschel, M.}} \yr{2014}  \at{{Ion
  temperature profile stiffness: non-linear gyrokinetic simulations and
  comparison with experiment}}.  \jt{Nuclear Fusion}  \bvol{54}~(2),
  \pg{023008}.

\bibitem[Colyer {\em et~al.\/}(2017)Colyer, Schekochihin, Parra, Roach, Barnes,
  Ghim \& Dorland]{Colyer2017}
{\sc \au{Colyer, G.~J.}, \au{Schekochihin, A.~A.}, \au{Parra, F.~I.},
  \au{Roach, C.~M.}, \au{Barnes, M.~A.}, \au{Ghim, Y.~C.} \& \au{Dorland, W.}}
  \yr{2017}  \at{{Collisionality scaling of the electron heat flux in ETG
  turbulence}}.  \jt{Plasma Physics and Controlled Fusion}  \bvol{59}~(5),
  \arxiv{arXiv: 1607.06752}.

\bibitem[Dimits {\em et~al.\/}(2000)Dimits, Bateman, Beer, Cohen, Dorland,
  Hammett, Kim, Kinsey, Kotschenreuther, Kritz, Lao, Mandrekas, Nevins, Parker,
  Redd, Shumaker, Sydora \& Weiland]{Dimits2000}
{\sc \au{Dimits, A.~M.}, \au{Bateman, G.}, \au{Beer, M.~A.}, \au{Cohen, B.~I.},
  \au{Dorland, W.}, \au{Hammett, G.~W.}, \au{Kim, C.}, \au{Kinsey, J.~E.},
  \au{Kotschenreuther, M.}, \au{Kritz, a.~H.}, \au{Lao, L.~L.}, \au{Mandrekas,
  J.}, \au{Nevins, W.~M.}, \au{Parker, S.~E.}, \au{Redd, a.~J.}, \au{Shumaker,
  D.~E.}, \au{Sydora, R.} \& \au{Weiland, J.}} \yr{2000}  \at{{Comparisons and
  physics basis of tokamak transport models and turbulence simulations}}.
  \jt{Physics of Plasmas}  \bvol{7}~(3),  \pg{969}.

\bibitem[Dorland \& Hammett(1993)]{Dorland1993a}
{\sc \au{Dorland, W.} \& \au{Hammett, G.~W.}} \yr{1993}  \at{{Gyrofluid
  Turbulence Models with Kinetic Effects}}.  \jt{Physics of Fluids B-Plasma
  Physics}  \bvol{5},  \pg{812--835}.

\bibitem[Dorland {\em et~al.\/}(2000)Dorland, Jenko, Kotschenreuther \&
  Rogers]{Dorland2000}
{\sc \au{Dorland, W.}, \au{Jenko, F.}, \au{Kotschenreuther, M.} \& \au{Rogers,
  B.~N.}} \yr{2000}  \at{{Electron temperature gradient turbulence.}}
  \jt{Physical review letters}  \bvol{85}~(26 Pt 1),  \pg{5579--5582}.

\bibitem[{EJ Doyle} {\em et~al.\/}(2007){EJ Doyle}, {WA Houlberg}, Kamada, {V
  Mukhovatov}, {TH Osborne}, {A Polevoi}, Bateman, {JW Connor}, {JG Cordey}, {T
  Fujita}, {X Garbet}, {TS Hahm}, {LD Horton}, {AE Hubbard}, {F Imbeaux}, {F
  Jenko}, {J. E. Kinsey}, {Y Kishimoto}, {J Li}, {TC Luce}, {Y Martin}, {M
  Ossipenko}, {V Parail}, {A Peeters}, {TL Rhodes}, {JE Rice}, Roach,
  Rozhansky, Ryter, {G Saibene}, {R Sartori}, {ACC Sips}, {JA Snipes}, {M
  Sugihara}, {EJ Synakowski}, {H Takenaga}, {T Takizuka}, {K Thomsen}, {MR
  Wade}, {HR Wilson}, {ITPA Transport Physics Topical Group}, {ITPA Confinement
  Database}, {Modelling Topical Group}, {ITPA Pedestal}, {Edge Topical Group},
  Physics), Database, Pedestal, Physics), Pedestal, Database, Bateman, Connor,
  (retired), Fujita, Garbet, Hahm, Horton, Hubbard, Imbeaux, Jenko, Kinsey,
  Kishimoto, Li, Luce, Martin, Ossipenko, Parail, Peeters, Rhodes, Rice, Roach,
  Rozhansky, Ryter, Saibene, Sartori, Sips, Snipes, Sugihara, Synakowski,
  Takenaga, Takizuka, Thomsen, Wade, Wilson, Group, Database, Group, Pedestal,
  Group, Doyle, Houlberg, Kamada, Mukhovatov, Osborne, Polevoi \&
  Cordey]{Doyle2007}
{\sc \au{{EJ Doyle}}, \au{{WA Houlberg}}, \au{Kamada, Y.}, \au{{V Mukhovatov}},
  \au{{TH Osborne}}, \au{{A Polevoi}}, \au{Bateman, G.}, \au{{JW Connor}},
  \au{{JG Cordey}}, \au{{T Fujita}}, \au{{X Garbet}}, \au{{TS Hahm}}, \au{{LD
  Horton}}, \au{{AE Hubbard}}, \au{{F Imbeaux}}, \au{{F Jenko}}, \au{{J. E.
  Kinsey}}, \au{{Y Kishimoto}}, \au{{J Li}}, \au{{TC Luce}}, \au{{Y Martin}},
  \au{{M Ossipenko}}, \au{{V Parail}}, \au{{A Peeters}}, \au{{TL Rhodes}},
  \au{{JE Rice}}, \au{Roach, C.~M.}, \au{Rozhansky, V.}, \au{Ryter, F.}, \au{{G
  Saibene}}, \au{{R Sartori}}, \au{{ACC Sips}}, \au{{JA Snipes}}, \au{{M
  Sugihara}}, \au{{EJ Synakowski}}, \au{{H Takenaga}}, \au{{T Takizuka}},
  \au{{K Thomsen}}, \au{{MR Wade}}, \au{{HR Wilson}}, \au{{ITPA Transport
  Physics Topical Group}}, \au{{ITPA Confinement Database}}, \au{{Modelling
  Topical Group}}, \au{{ITPA Pedestal}}, \au{{Edge Topical Group}},
  \au{Physics), E. J. D. C.~T.}, \au{Database, W. A. H. C.~C.}, \au{Pedestal,
  Y. K.~C.}, \au{Physics), V. M. c.-C.~T.}, \au{Pedestal, T. H. O. c.-C.},
  \au{Database, A. P. c.-C.~C.}, \au{Bateman, G.}, \au{Connor, J.~W.},
  \au{(retired), J. G.~C.}, \au{Fujita, T.}, \au{Garbet, X.}, \au{Hahm, T.~S.},
  \au{Horton, L.~D.}, \au{Hubbard, A.~E.}, \au{Imbeaux, F.}, \au{Jenko, F.},
  \au{Kinsey, J.~E.}, \au{Kishimoto, Y.}, \au{Li, J.}, \au{Luce, T.~C.},
  \au{Martin, Y.}, \au{Ossipenko, M.}, \au{Parail, V.}, \au{Peeters, A.},
  \au{Rhodes, T.~L.}, \au{Rice, J.~E.}, \au{Roach, C.~M.}, \au{Rozhansky, V.},
  \au{Ryter, F.}, \au{Saibene, G.}, \au{Sartori, R.}, \au{Sips, A. C.~C.},
  \au{Snipes, J.~A.}, \au{Sugihara, M.}, \au{Synakowski, E.~J.}, \au{Takenaga,
  H.}, \au{Takizuka, T.}, \au{Thomsen, K.}, \au{Wade, M.~R.}, \au{Wilson,
  H.~R.}, \au{Group, I. T. P.~T.}, \au{Database, I.~C.}, \au{Group, M.~T.},
  \au{Pedestal, I.}, \au{Group, E.~T.}, \au{Doyle, E.~J.}, \au{Houlberg,
  W.~A.}, \au{Kamada, Y.}, \au{Mukhovatov, V.}, \au{Osborne, T.~H.},
  \au{Polevoi, A.} \& \au{Cordey, J.~G.}} \yr{2007}  \at{{Chapter 2: Plasma
  confinement and transport}}.  \jt{Nuclear Fusion}  \bvol{47}~(6),  \pg{S18}.

\bibitem[ESTECO(2018)]{modefrontier}
{\sc \au{ESTECO, M.}} \yr{2018} {modeFrontier Home Page}.
  http://www.esteco.com/modefrontier.

\bibitem[Federici {\em et~al.\/}(2014)Federici, Kemp, Ward, Bachmann, Franke,
  Gonzalez, Lowry, Gadomska, Harman, Meszaros, Morlock, Romanelli \&
  Wenninger]{Federici2014}
{\sc \au{Federici, G.}, \au{Kemp, R.}, \au{Ward, D.}, \au{Bachmann, C.},
  \au{Franke, T.}, \au{Gonzalez, S.}, \au{Lowry, C.}, \au{Gadomska, M.},
  \au{Harman, J.}, \au{Meszaros, B.}, \au{Morlock, C.}, \au{Romanelli, F.} \&
  \au{Wenninger, R.}} \yr{2014}  \at{{Overview of EU DEMO design and R{\&}D
  activities}}.  \jt{Fusion Engineering and Design}  \bvol{89}~(7-8),
  \pg{882--889}.

\bibitem[Frieman \& Chen(1982)]{Frieman1982}
{\sc \au{Frieman, E.~A.} \& \au{Chen, L.}} \yr{1982}  \at{{Nonlinear
  gyrokinetic equations for low-frequency electromagnetic waves in general
  plasma equilibria}}.  \jt{Physics of Fluids}  \bvol{25}~(3),  \pg{502}.

\bibitem[Galambos {\em et~al.\/}(1995)Galambos, Perkins, Haney \&
  Mandrekas]{Galambos1995}
{\sc \au{Galambos, J.}, \au{Perkins, L.}, \au{Haney, S.} \& \au{Mandrekas, J.}}
  \yr{1995}  \at{{Commercial tokamak reactor potential with advanced tokamak
  operation}}.  \jt{Nuclear Fusion}  \bvol{35}~(5),  \pg{551--573}.

\bibitem[Glasstone \& Lovberg(1960)]{Glasstone1960}
{\sc \au{Glasstone, S.} \& \au{Lovberg, R.~H.}} \yr{1960} {Controlled
  Thermonuclear Reactions. D. Van Noatrand Company}.

\bibitem[Hammett \& Perkins(1990)]{Hammett1990}
{\sc \au{Hammett, G.~W.} \& \au{Perkins, F.~W.}} \yr{1990}  \at{{Fluid moment
  models for Landau damping with application to the ion-temperature-gradient
  instability}}.  \jt{Physical review letters}  \bvol{64}~(25),
  \pg{3019--3022}.

\bibitem[Highcock {\em et~al.\/}(2011)Highcock, Barnes, Parra, Schekochihin,
  Roach \& Cowley]{Highcock2011}
{\sc \au{Highcock, E.~G.}, \au{Barnes, M.}, \au{Parra, F.~I.},
  \au{Schekochihin, A.~A.}, \au{Roach, C.~M.} \& \au{Cowley, S.~C.}} \yr{2011}
  \at{{Transport bifurcation induced by sheared toroidal flow in tokamak
  plasmas}}.  \jt{Physics of Plasmas}  \bvol{18}~(10),  \pg{102304}.

\bibitem[Highcock {\em et~al.\/}(2010)Highcock, Barnes, Schekochihin, Parra,
  Roach \& Cowley]{Highcock2010}
{\sc \au{Highcock, E.~G.}, \au{Barnes, M.}, \au{Schekochihin, A.~A.},
  \au{Parra, F.~I.}, \au{Roach, C.~M.} \& \au{Cowley, S.~C.}} \yr{2010}
  \at{{Transport Bifurcation in a Rotating Tokamak Plasma}}.  \jt{Physical
  Review Letters}  \bvol{105}~(21),  \pg{215003}.

\bibitem[Highcock {\em et~al.\/}(2012)Highcock, Schekochihin, Cowley, Barnes,
  Parra, Roach \& Dorland]{Highcock2012}
{\sc \au{Highcock, E.~G.}, \au{Schekochihin, A.~A.}, \au{Cowley, S.~C.},
  \au{Barnes, M.}, \au{Parra, F.~I.}, \au{Roach, C.~M.} \& \au{Dorland, W.}}
  \yr{2012}  \at{{Zero-Turbulence Manifold in a Toroidal Plasma}}.
  \jt{Physical Review Letters}  \bvol{109}~(26),  \pg{265001}.

\bibitem[Howes {\em et~al.\/}(2011)Howes, TenBarge, Dorland, Quataert,
  Schekochihin, Numata \& Tatsuno]{Howes2011}
{\sc \au{Howes, G.~G.}, \au{TenBarge, J.~M.}, \au{Dorland, W.}, \au{Quataert,
  E.}, \au{Schekochihin, A.~A.}, \au{Numata, R.} \& \au{Tatsuno, T.}} \yr{2011}
   \at{Gyrokinetic simulations of solar wind turbulence from ion to electron
  scales}.  \jt{Physical review letters}  \bvol{107}~(3),  \pg{035004}.

\bibitem[Jardin {\em et~al.\/}(2006)Jardin, Kessel, Mau, Miller, Najmabadi,
  Chan, Chu, Lahaye, Lao, Petrie, Politzer, {St. John}, Snyder, Staebler,
  Turnbull \& West]{Jardin2006}
{\sc \au{Jardin, S.~C.}, \au{Kessel, C.~E.}, \au{Mau, T.~K.}, \au{Miller,
  R.~L.}, \au{Najmabadi, F.}, \au{Chan, V.~S.}, \au{Chu, M.~S.}, \au{Lahaye,
  R.}, \au{Lao, L.~L.}, \au{Petrie, T.~W.}, \au{Politzer, P.}, \au{{St. John},
  H.~E.}, \au{Snyder, P.}, \au{Staebler, G.~M.}, \au{Turnbull, A.~D.} \&
  \au{West, W.~P.}} \yr{2006}  \at{{Physics basis for the advanced tokamak
  fusion power plant, ARIES-AT}}.  \jt{Fusion Engineering and Design}
  \bvol{80}~(1-4),  \pg{25--62}.

\bibitem[Kawamori(2013)]{Kawamori2013}
{\sc \au{Kawamori, E.}} \yr{2013}  \at{Experimental verification of entropy
  cascade in two-dimensional electrostatic turbulence in magnetized plasma}.
  \jt{Physical review letters}  \bvol{110}~(9),  \pg{095001}.

\bibitem[Kinsey {\em et~al.\/}(2011)Kinsey, Staebler, Candy, Waltz \&
  Budny]{Kinsey2011}
{\sc \au{Kinsey, J.~E.}, \au{Staebler, G.~M.}, \au{Candy, J.}, \au{Waltz,
  R.~E.} \& \au{Budny, R.~V.}} \yr{2011}  \at{{ITER predictions using the GYRO
  verified and experimentally validated trapped gyro-Landau fluid transport
  model}}.  \jt{Nuclear Fusion}  \bvol{51}~(8),  \pg{083001}.

\bibitem[Kotschenreuther {\em et~al.\/}(1995)Kotschenreuther, Rewoldt \&
  Tang]{Kotschenreuther1995c}
{\sc \au{Kotschenreuther, M.}, \au{Rewoldt, G.} \& \au{Tang, W.~M.}} \yr{1995}
  \at{{Comparison of initial value and eigenvalue codes for kinetic toroidal
  plasma instabilities}}.  \jt{Computer Physics Communications}
  \bvol{88}~(2-3),  \pg{128--140}.

\bibitem[Lewis(2004)]{Lewis2004b}
{\sc \au{Lewis, A.}} \yr{2004}  \at{{Parallel Optimisation Algorithms for
  Continuous, Non-Linear Numerical Simulations}}. PhD thesis, Griffith
  University, Brisbane, Australia.

\bibitem[Luce {\em et~al.\/}(2014)Luce, Challis, Ide, Joffrin, Kamada,
  Politzer, Schweinzer, a.C.C. Sips, Stober, Giruzzi, Kessel, Murakami, Na,
  Park, a.R. Polevoi, Budny, Citrin, Garcia, Hayashi, Hobirk, Hudson, Imbeaux,
  Isayama, McDonald, Nakano, Oyama, Parail, Petrie, Petty, Suzuki \&
  Wade]{Luce2014}
{\sc \au{Luce, T.}, \au{Challis, C.}, \au{Ide, S.}, \au{Joffrin, E.},
  \au{Kamada, Y.}, \au{Politzer, P.}, \au{Schweinzer, J.}, \au{a.C.C. Sips},
  \au{Stober, J.}, \au{Giruzzi, G.}, \au{Kessel, C.}, \au{Murakami, M.},
  \au{Na, Y.-S.}, \au{Park, J.}, \au{a.R. Polevoi}, \au{Budny, R.}, \au{Citrin,
  J.}, \au{Garcia, J.}, \au{Hayashi, N.}, \au{Hobirk, J.}, \au{Hudson, B.},
  \au{Imbeaux, F.}, \au{Isayama, a.}, \au{McDonald, D.}, \au{Nakano, T.},
  \au{Oyama, N.}, \au{Parail, V.}, \au{Petrie, T.}, \au{Petty, C.}, \au{Suzuki,
  T.} \& \au{Wade, M.}} \yr{2014}  \at{{Development of advanced inductive
  scenarios for ITER}}.  \jt{Nuclear Fusion}  \bvol{54}~(1),  \pg{013015}.

\bibitem[L{\"{u}}tjens {\em et~al.\/}(1996)L{\"{u}}tjens, Bondeson \&
  Sauter]{Bondeson1996}
{\sc \au{L{\"{u}}tjens, H.}, \au{Bondeson, A.} \& \au{Sauter, O.}} \yr{1996}
  \at{{The CHEASE code for toroidal MHD equilibria}}.  \jt{Computer physics
  communications}  \bvol{97},  \pg{219--260}.

\bibitem[Maggi {\em et~al.\/}(2007)Maggi, Groebner, Oyama, Sartori, Horton,
  Sips, Suttrop, Leonard, Luce, Wade, Kamada, Urano, Andrew, Giroud, Joffrin \&
  {De La Luna}]{Maggi2007a}
{\sc \au{Maggi, C.~F.}, \au{Groebner, R.~J.}, \au{Oyama, N.}, \au{Sartori, R.},
  \au{Horton, L.~D.}, \au{Sips, A.~C.}, \au{Suttrop, W.}, \au{Leonard, A.},
  \au{Luce, T.~C.}, \au{Wade, M.~R.}, \au{Kamada, Y.}, \au{Urano, H.},
  \au{Andrew, Y.}, \au{Giroud, C.}, \au{Joffrin, E.} \& \au{{De La Luna}, E.}}
  \yr{2007}  \at{{Characteristics of the H-mode pedestal in improved
  confinement scenarios in ASDEX upgrade, DIII-D, JET and JT-60U}}.
  \jt{Nuclear Fusion}  \bvol{47}~(7),  \pg{535--551}.

\bibitem[Mandell \& Dorland(2014)]{Mandell2014}
{\sc \au{Mandell, N.} \& \au{Dorland, W.}} \yr{2014} {Hybrid
  Gyrokinetic/Gyrofluid Simulation of ITG Turbulence}.  \bt{In {\em Bulletin of
  the American Physical Society\/}},  \pg{p. Abstract CP8.039}.

\bibitem[Mandell {\em et~al.\/}(2018)Mandell, Dorland \&
  Landreman]{Mandell2018}
{\sc \au{Mandell, N.}, \au{Dorland, W.} \& \au{Landreman, M.}} \yr{2018}
  \at{Laguerre--hermite pseudo-spectral velocity formulation of gyrokinetics}.
  \jt{Journal of Plasma Physics}  \bvol{84}~(1).

\bibitem[Manousopoulos \& Michalopoulos(2009)]{Manousopoulos2009}
{\sc \au{Manousopoulos, P.} \& \au{Michalopoulos, M.}} \yr{2009}
  \at{{Comparison of non-linear optimization algorithms for yield curve
  estimation}}.  \jt{European Journal of Operational Research}  \bvol{192}~(2),
   \pg{594--602}.

\bibitem[Marinoni {\em et~al.\/}(2009)Marinoni, Brunner, Camenen, Coda, Graves,
  Lapillonne, Pochelon, Sauter \& Villard]{Marinoni2009}
{\sc \au{Marinoni, A.}, \au{Brunner, S.}, \au{Camenen, Y.}, \au{Coda, S.},
  \au{Graves, J.~P.}, \au{Lapillonne, X.}, \au{Pochelon, A.}, \au{Sauter, O.}
  \& \au{Villard, L.}} \yr{2009}  \at{{The effect of plasma triangularity on
  turbulent transport: modeling TCV experiments by linear and non-linear
  gyrokinetic simulations}}.  \jt{Plasma Physics and Controlled Fusion}
  \bvol{51}~(5),  \pg{055016}.

\bibitem[Meneghini {\em et~al.\/}(2016)Meneghini, Snyder, Smith, Candy,
  Staebler, Belli, Lao, Park, Green, Elwasif, Grierson \&
  Holland]{Meneghini2016}
{\sc \au{Meneghini, O.}, \au{Snyder, P.~B.}, \au{Smith, S.~P.}, \au{Candy, J.},
  \au{Staebler, G.~M.}, \au{Belli, E.~A.}, \au{Lao, L.~L.}, \au{Park, J.~M.},
  \au{Green, D.~L.}, \au{Elwasif, W.}, \au{Grierson, B.~A.} \& \au{Holland,
  C.}} \yr{2016}  \at{{Integrated fusion simulation with self-consistent
  core-pedestal coupling}}.  \jt{Physics of Plasmas}  \bvol{23}~(4),
  \pg{042507}.

\bibitem[Merle {\em et~al.\/}(2017)Merle, Sauter \& {Yu Medvedev}]{Merle2017}
{\sc \au{Merle, A.}, \au{Sauter, O.} \& \au{{Yu Medvedev}, S.}} \yr{2017}
  \at{{Pedestal properties of H-modes with negative triangularity using the
  EPED-CH model}}.  \jt{Plasma Physics and Controlled Fusion}  \bvol{59}~(10).

\bibitem[Mukhovatov {\em et~al.\/}(2003)Mukhovatov, Shimomura, Polevoi,
  Shimada, Sugihara, Bateman, Cordey, Kardaun, Pereverzev, Voitsekhovitch,
  Weiland, Zolotukhin, Chudnovskiy, Kritz, Kukushkin, Onjun, Pankin \&
  Perkins]{Mukhovatov2003}
{\sc \au{Mukhovatov, V.}, \au{Shimomura, Y.}, \au{Polevoi, A.}, \au{Shimada,
  M.}, \au{Sugihara, M.}, \au{Bateman, G.}, \au{Cordey, J.~G.}, \au{Kardaun,
  O.}, \au{Pereverzev, G.}, \au{Voitsekhovitch, I.}, \au{Weiland, J.},
  \au{Zolotukhin, O.}, \au{Chudnovskiy, A.}, \au{Kritz, A.~H.}, \au{Kukushkin,
  A.}, \au{Onjun, T.}, \au{Pankin, A.} \& \au{Perkins, F.~W.}} \yr{2003}
  \at{{Comparison of ITER performance predicted by semi-empirical and
  theory-based transport models}}.  \jt{Nuclear Fusion}  \bvol{43}~(9),
  \pg{942--948}.

\bibitem[Parail {\em et~al.\/}(2013)Parail, Albanese, Ambrosino, Artaud,
  Besseghir, Cavinato, Corrigan, Garcia, Garzotti, Gribov, Imbeaux, Koechl,
  Labate, Lister, Litaudon, Loarte, Maget, Mattei, McDonald, Nardon, Saibene,
  Sartori \& Urban]{Parail2013}
{\sc \au{Parail, V.}, \au{Albanese, R.}, \au{Ambrosino, R.}, \au{Artaud,
  J.~F.}, \au{Besseghir, K.}, \au{Cavinato, M.}, \au{Corrigan, G.}, \au{Garcia,
  J.}, \au{Garzotti, L.}, \au{Gribov, Y.}, \au{Imbeaux, F.}, \au{Koechl, F.},
  \au{Labate, C.~V.}, \au{Lister, J.}, \au{Litaudon, X.}, \au{Loarte, A.},
  \au{Maget, P.}, \au{Mattei, M.}, \au{McDonald, D.}, \au{Nardon, E.},
  \au{Saibene, G.}, \au{Sartori, R.} \& \au{Urban, J.}} \yr{2013}
  \at{{Self-consistent simulation of plasma scenarios for ITER using a
  combination of 1.5D transport codes and free-boundary equilibrium codes}}.
  \jt{Nuclear Fusion}  \bvol{53}~(11),  \pg{113002}.

\bibitem[Parker {\em et~al.\/}(2016)Parker, Highcock, Schekochihin \&
  Dellar]{Parker2016a}
{\sc \au{Parker, J.~T.}, \au{Highcock, E.~G.}, \au{Schekochihin, A.~A.} \&
  \au{Dellar, P.~J.}} \yr{2016}  \at{{Suppression of phase mixing in
  drift-kinetic plasma turbulence}}.  \jt{Physics of Plasmas}  \bvol{23}~(7),
  \pg{070703},  \arxiv{arXiv: 1603.06968}.

\bibitem[Pochelon {\em et~al.\/}(2012)Pochelon, Angelino, Behn, Brunner, Coda,
  Kirneva, Medvedev, Reimerdes, Rossel, Sauter, Villard, W{\'{a}}gner, Bottino,
  Camenen, Canal, Chattopadhyay, Duval, Fasoli, Goodman, Jolliet, Karpushov,
  Labit, Marinoni, Moret, Pitzschke, Porte, Rancic \& Udintsev]{Pochelon2012}
{\sc \au{Pochelon, A.}, \au{Angelino, P.}, \au{Behn, R.}, \au{Brunner, S.},
  \au{Coda, S.}, \au{Kirneva, N.}, \au{Medvedev, S.}, \au{Reimerdes, H.},
  \au{Rossel, J.}, \au{Sauter, O.}, \au{Villard, L.}, \au{W{\'{a}}gner, D.},
  \au{Bottino, A.}, \au{Camenen, Y.}, \au{Canal, G.~P.}, \au{Chattopadhyay,
  P.~K.}, \au{Duval, B.~P.}, \au{Fasoli, A.}, \au{Goodman, T.~P.}, \au{Jolliet,
  S.}, \au{Karpushov, A.}, \au{Labit, B.}, \au{Marinoni, A.}, \au{Moret,
  J.~M.}, \au{Pitzschke, A.}, \au{Porte, L.}, \au{Rancic, M.} \& \au{Udintsev,
  V.~S.}} \yr{2012}  \at{{Recent TCV results - Innovative plasma shaping to
  improve plasma properties and insight}}.  \jt{Plasma and Fusion Research}
  \bvol{7}~(SPL.ISS.1),  \pg{1--8}.

\bibitem[Pochelon {\em et~al.\/}(1999)Pochelon, Goodman, Henderson, Angioni,
  Behn, Coda, Hofmann, Hogge, Kirneva, Martynov, Moret, Pietrzyk, Porcelli,
  Reimerdes, Rommers, Rossi, Sauter, Tran, Weisen, Alberti, Barry, Blanchard,
  Bosshard, Chavan, Duval, Esipchuck, Fasel, Favre, Franke, Furno, Gorgerat,
  Isoz, Joye, Lister, Llobet, Magnin, Mandrin, Manini, Marl{\'{e}}taz,
  Marmillod, Martin, Mayor, Mlynar, Nieswand, Paris, Perez, Pitts, Razumova,
  Refke, Scavino, Sushkov, Tonetti, Troyon, Toledo \& Vyas]{Pochelon1999}
{\sc \au{Pochelon, A.}, \au{Goodman, T.}, \au{Henderson, M.}, \au{Angioni, C.},
  \au{Behn, R.}, \au{Coda, S.}, \au{Hofmann, F.}, \au{Hogge, J.-P.},
  \au{Kirneva, N.}, \au{Martynov, A.}, \au{Moret, J.-M.}, \au{Pietrzyk, Z.~A.},
  \au{Porcelli, F.}, \au{Reimerdes, H.}, \au{Rommers, J.}, \au{Rossi, E.},
  \au{Sauter, O.}, \au{Tran, M.~Q.}, \au{Weisen, H.}, \au{Alberti, S.},
  \au{Barry, S.}, \au{Blanchard, P.}, \au{Bosshard, P.}, \au{Chavan, R.},
  \au{Duval, B.~P.}, \au{Esipchuck, Y.~V.}, \au{Fasel, D.}, \au{Favre, A.},
  \au{Franke, S.}, \au{Furno, I.}, \au{Gorgerat, P.}, \au{Isoz, P.-F.},
  \au{Joye, B.}, \au{Lister, J.}, \au{Llobet, X.}, \au{Magnin, J.-C.},
  \au{Mandrin, P.}, \au{Manini, A.}, \au{Marl{\'{e}}taz, B.}, \au{Marmillod,
  P.}, \au{Martin, Y.}, \au{Mayor, J.-M.}, \au{Mlynar, J.}, \au{Nieswand, C.},
  \au{Paris, P.}, \au{Perez, A.}, \au{Pitts, R.}, \au{Razumova, K.}, \au{Refke,
  A.}, \au{Scavino, E.}, \au{Sushkov, A.}, \au{Tonetti, G.}, \au{Troyon, F.},
  \au{Toledo, W.~V.} \& \au{Vyas, P.}} \yr{1999}  \at{{Energy confinement and
  MHD activity in shaped TCV plasmas with localized electron cyclotron
  heating}}.  \jt{Nuclear Fusion}  \bvol{39}~(11Y),  \pg{1807--1818}.

\bibitem[Roach {\em et~al.\/}(2008)Roach, Walters, Budny, Imbeaux, Fredian,
  Greenwald, Stillerman, Alexander, Carlsson, Cary, Ryter, Stober, Gohil,
  Greenfield, Murakami, Bracco, Esposito, Romanelli, Parail, Stubberfield,
  Voitsekhovitch, Brickley, Field, Sakamoto, Fujita, Fukuda, Hayashi, Hogeweij,
  Chudnovskiy, Kinerva, Kessel, Aniel, Hoang, Ongena, Doyle, Houlberg \&
  Polevoi]{Roach2008}
{\sc \au{Roach, C.}, \au{Walters, M.}, \au{Budny, R.}, \au{Imbeaux, F.},
  \au{Fredian, T.}, \au{Greenwald, M.}, \au{Stillerman, J.}, \au{Alexander,
  D.}, \au{Carlsson, J.}, \au{Cary, J.}, \au{Ryter, F.}, \au{Stober, J.},
  \au{Gohil, P.}, \au{Greenfield, C.}, \au{Murakami, M.}, \au{Bracco, G.},
  \au{Esposito, B.}, \au{Romanelli, M.}, \au{Parail, V.}, \au{Stubberfield,
  P.}, \au{Voitsekhovitch, I.}, \au{Brickley, C.}, \au{Field, A.},
  \au{Sakamoto, Y.}, \au{Fujita, T.}, \au{Fukuda, T.}, \au{Hayashi, N.},
  \au{Hogeweij, G.}, \au{Chudnovskiy, A.}, \au{Kinerva, N.}, \au{Kessel, C.},
  \au{Aniel, T.}, \au{Hoang, G.}, \au{Ongena, J.}, \au{Doyle, E.},
  \au{Houlberg, W.} \& \au{Polevoi, A.}} \yr{2008}  \at{{The 2008 Public
  Release of the International Multi-tokamak Confinement Profile Database}}.
  \jt{Nuclear Fusion}  \bvol{48}~(12),  \pg{125001}.

\bibitem[Rogers {\em et~al.\/}(2000)Rogers, Dorland \&
  Kotschenreuther]{Rogers2000}
{\sc \au{Rogers, B.~N.}, \au{Dorland, W.} \& \au{Kotschenreuther, M.}}
  \yr{2000}  \at{{Generation and stability of zonal flows in
  ion-temperature-gradient mode turbulence.}}  \jt{Physical review letters}
  \bvol{85}~(25),  \pg{5336--9}.

\bibitem[Rosenbluth \& Hinton(1998)]{Rosenbluth1998}
{\sc \au{Rosenbluth, M.} \& \au{Hinton, F.}} \yr{1998}  \at{Poloidal flow
  driven by ion-temperature-gradient turbulence in tokamaks}.  \jt{Physical
  review letters}  \bvol{80}~(4),  \pg{724}.

\bibitem[Schekochihin {\em et~al.\/}(2009)Schekochihin, Cowley, Dorland,
  Hammett, Howes, Quataert \& Tatsuno]{Schekochihin2009}
{\sc \au{Schekochihin, A.~A.}, \au{Cowley, S.~C.}, \au{Dorland, W.},
  \au{Hammett, G.~W.}, \au{Howes, G.~G.}, \au{Quataert, E.} \& \au{Tatsuno,
  T.}} \yr{2009}  \at{{Astrophysical Gyrokinetics: Kinetic and Fluid Turbulent
  Cascades in Magnetized Weakly Collisional Plasmas}}.  \jt{The Astrophysical
  Journal Supplement Series}  \bvol{182}~(1),  \pg{310--377}.

\bibitem[Schekochihin {\em et~al.\/}(2016)Schekochihin, Parker, Highcock,
  Dellar, Dorland \& Hammett]{Schekochihin2016}
{\sc \au{Schekochihin, A.~A.}, \au{Parker, J.~T.}, \au{Highcock, E.~G.},
  \au{Dellar, P.~J.}, \au{Dorland, W.} \& \au{Hammett, G.~W.}} \yr{2016}
  \at{{Phase mixing versus nonlinear advection in drift-kinetic plasma
  turbulence}}.  \jt{Journal of Plasma Physics}  \bvol{82}~(02),
  \pg{905820212},  \arxiv{arXiv: 1508.05988}.

\bibitem[Simpson {\em et~al.\/}(2008)Simpson, Toropov, Balabanov \&
  Viana]{Simpson2008}
{\sc \au{Simpson, T.}, \au{Toropov, V.}, \au{Balabanov, V.} \& \au{Viana, F.}}
  \yr{2008}  \at{{Design and Analysis of Computer Experiments in
  Multidisciplinary Design Optimization: A Review of How Far We Have Come - Or
  Not}}.  \jt{12th AIAA/ISSMO Multidisciplinary Analysis and Optimization
  Conference} ~(September),  \pg{1--22}.

\bibitem[Snyder \& Hammett(2001)]{Snyder2001}
{\sc \au{Snyder, P.} \& \au{Hammett, G.}} \yr{2001}  \at{A landau fluid model
  for electromagnetic plasma microturbulence}.  \jt{Physics of Plasmas}
  \bvol{8}~(7),  \pg{3199--3216}.

\bibitem[Sorbom {\em et~al.\/}(2015)Sorbom, Ball, Palmer, Mangiarotti,
  Sierchio, Bonoli, Kasten, Sutherland, Barnard, Haakonsen, Goh, Sung \&
  Whyte]{Sorbom2015}
{\sc \au{Sorbom, B.~N.}, \au{Ball, J.}, \au{Palmer, T.~R.}, \au{Mangiarotti,
  F.~J.}, \au{Sierchio, J.~M.}, \au{Bonoli, P.}, \au{Kasten, C.},
  \au{Sutherland, D.~A.}, \au{Barnard, H.~S.}, \au{Haakonsen, C.~B.}, \au{Goh,
  J.}, \au{Sung, C.} \& \au{Whyte, D.~G.}} \yr{2015}  \at{{ARC: A compact,
  high-field, fusion nuclear science facility and demonstration power plant
  with demountable magnets}}.  \jt{Fusion Engineering and Design}  \bvol{100},
  \pg{378--405},  \arxiv{arXiv: 1409.3540}.

\bibitem[Staebler \& John(2006)]{Staebler2006}
{\sc \au{Staebler, G.~M.} \& \au{John, H. E.~S.}} \yr{2006}  \at{{Predicted
  toroidal rotation enhancement of fusion power production in ITER}}.
  \jt{Nuclear Fusion}  \bvol{46}~(8),  \pg{L6--L8}.

\bibitem[Staebler {\em et~al.\/}(2007)Staebler, Kinsey \& Waltz]{Staebler2007}
{\sc \au{Staebler, G.~M.}, \au{Kinsey, J.~E.} \& \au{Waltz, R.~E.}} \yr{2007}
  \at{{A theory-based transport model with comprehensive physics}}.
  \jt{Physics of Plasmas}  \bvol{14}~(5).

\bibitem[Stork {\em et~al.\/}(2014)Stork, Agostini, Boutard, Buckthorpe,
  Diegele, Dudarev, English, Federici, Gilbert, Gonzalez, Ibarra, Linsmeier,
  Puma, Marbach, Packer, Raj, Rieth, Tran, Ward \& Zinkle]{Stork2014}
{\sc \au{Stork, D.}, \au{Agostini, P.}, \au{Boutard, J.~L.}, \au{Buckthorpe,
  D.}, \au{Diegele, E.}, \au{Dudarev, S.~L.}, \au{English, C.}, \au{Federici,
  G.}, \au{Gilbert, M.~R.}, \au{Gonzalez, S.}, \au{Ibarra, A.}, \au{Linsmeier,
  C.}, \au{Puma, A.~L.}, \au{Marbach, G.}, \au{Packer, L.~W.}, \au{Raj, B.},
  \au{Rieth, M.}, \au{Tran, M.~Q.}, \au{Ward, D.~J.} \& \au{Zinkle, S.~J.}}
  \yr{2014}  \at{{Materials R{\&}D for a timely DEMO: Key findings and
  recommendations of the EU Roadmap Materials Assessment Group}}.  \jt{Fusion
  Engineering and Design}  \bvol{89}~(7-8),  \pg{1586--1594}.

\bibitem[Sugama \& Horton(1998)]{Sugama1998}
{\sc \au{Sugama, H.} \& \au{Horton, W.}} \yr{1998}  \at{{Nonlinear
  electromagnetic gyrokinetic equation for plasmas with large mean flows}}.
  \jt{Physics of Plasmas}  \bvol{5}~(7),  \pg{2560}.

\bibitem[Synakowski(1999)]{Synakowski1999}
{\sc \au{Synakowski, E.~J.}} \yr{1999}  \at{{Formation and structure of
  internal and edge transport barriers}}.  \jt{Plasma Physics and Controlled
  Fusion}  \bvol{40}~(5),  \pg{581--596}.

\bibitem[Tatsuno {\em et~al.\/}(2009)Tatsuno, Dorland, Schekochihin, Plunk,
  Barnes, Cowley \& Howes]{Tatsuno2009}
{\sc \au{Tatsuno, T.}, \au{Dorland, W.}, \au{Schekochihin, A.~A.}, \au{Plunk,
  G.~G.}, \au{Barnes, M.}, \au{Cowley, S.~C.} \& \au{Howes, G.~G.}} \yr{2009}
  \at{{Nonlinear phase mixing and phase-space cascade of entropy in gyrokinetic
  plasma turbulence}}.  \jt{Physical Review Letters}  \bvol{103}~(1),
  \pg{2--5},  \arxiv{arXiv: 0811.2538}.

\bibitem[Tatsuno {\em et~al.\/}(2012)Tatsuno, Plunk, Barnes, Dorland, Howes \&
  Numata]{Tatsuno2012}
{\sc \au{Tatsuno, T.}, \au{Plunk, G.}, \au{Barnes, M.}, \au{Dorland, W.},
  \au{Howes, G.} \& \au{Numata, R.}} \yr{2012}  \at{Freely decaying turbulence
  in two-dimensional electrostatic gyrokinetics}.  \jt{Physics of Plasmas}
  \bvol{19}~(12),  \pg{122305}.

\bibitem[Uckan(1990)]{Uckan1990}
{\sc \au{Uckan, N.}} \yr{1990} {ITER physics design guidelines: 1989}.  \bt{In
  {\em ITER Documentation Series 10 (Vienna: IAEA)\/}}.

\bibitem[Wagner {\em et~al.\/}(1984)Wagner, Fussmann, Grave, Keilhacker,
  Kornherr, Lackner, McCormick, M{\"{u}}ller, St{\"{a}}bler, Becker, Bernhardi,
  Ditte, Eberhagen, Gehre, Gernhardt, Gierke, Glock, Gruber, Haas, Hesse,
  Janeschitz, Karger, Kissel, Kl{\"{u}}ber, Lisitano, Mayer, Meisel, Mertens,
  Murmann, Poschenrieder, Rapp, R{\"{o}}hr, Ryter, Schneider, Siller,
  Smeulders, S{\"{o}}ldner, Speth, Steuer, Szymanski \& Vollmer]{Wagner1984}
{\sc \au{Wagner, F.}, \au{Fussmann, G.}, \au{Grave, T.}, \au{Keilhacker, M.},
  \au{Kornherr, M.}, \au{Lackner, K.}, \au{McCormick, K.}, \au{M{\"{u}}ller,
  E.~R.}, \au{St{\"{a}}bler, A.}, \au{Becker, G.}, \au{Bernhardi, K.},
  \au{Ditte, U.}, \au{Eberhagen, A.}, \au{Gehre, O.}, \au{Gernhardt, J.},
  \au{Gierke, G.~v.}, \au{Glock, E.}, \au{Gruber, O.}, \au{Haas, G.},
  \au{Hesse, M.}, \au{Janeschitz, G.}, \au{Karger, F.}, \au{Kissel, S.},
  \au{Kl{\"{u}}ber, O.}, \au{Lisitano, G.}, \au{Mayer, H.~M.}, \au{Meisel, D.},
  \au{Mertens, V.}, \au{Murmann, H.}, \au{Poschenrieder, W.}, \au{Rapp, H.},
  \au{R{\"{o}}hr, H.}, \au{Ryter, F.}, \au{Schneider, F.}, \au{Siller, G.},
  \au{Smeulders, P.}, \au{S{\"{o}}ldner, F.}, \au{Speth, E.}, \au{Steuer,
  K.~H.}, \au{Szymanski, Z.} \& \au{Vollmer, O.}} \yr{1984}  \at{{Development
  of an Edge Transport Barrier at the H-Mode Transition of ASDEX}}.
  \jt{Physical Review Letters}  \bvol{53}~(15),  \pg{1453--1456}.

\bibitem[Weisen {\em et~al.\/}(1999)Weisen, Alberti, Berry, Behn, Blanchard,
  Bosshard, B{\"{u}}hlmann, Chavan, Coda, Deschenaux, Dutch, Duval, Fasel,
  Favre, Franke, Furno, Goodman, Henderson, Hofmann, Hogge, Isoz, Joye, Lister,
  Llobet, Magnin, Mandrin, Marletaz, Marmillod, Martin, Mayor, Moret, Nieswand,
  Paris, Perez, Pietrzyk, Piffl, Pitts, Pochelon, Razumova, Reimerdes, Refke,
  Rommers, Roy, Sauter, Suttrop, Toledo, Tonetti, Tran, Troyon, Vyas \&
  Ward]{Weisen1999}
{\sc \au{Weisen, H.}, \au{Alberti, S.}, \au{Berry, S.}, \au{Behn, R.},
  \au{Blanchard, P.}, \au{Bosshard, P.}, \au{B{\"{u}}hlmann, F.}, \au{Chavan,
  R.}, \au{Coda, S.}, \au{Deschenaux, C.}, \au{Dutch, M.~J.}, \au{Duval,
  B.~P.}, \au{Fasel, D.}, \au{Favre, A.}, \au{Franke, S.}, \au{Furno, I.},
  \au{Goodman, T.}, \au{Henderson, M.}, \au{Hofmann, F.}, \au{Hogge, J.-P.},
  \au{Isoz, P.-F.}, \au{Joye, B.}, \au{Lister, J.~B.}, \au{Llobet, X.},
  \au{Magnin, J.-C.}, \au{Mandrin, P.}, \au{Marletaz, B.}, \au{Marmillod, P.},
  \au{Martin, Y.}, \au{Mayor, J.-M.}, \au{Moret, J.-M.}, \au{Nieswand, C.},
  \au{Paris, P.}, \au{Perez, A.}, \au{Pietrzyk, Z.~a.}, \au{Piffl, V.},
  \au{Pitts, R.~a.}, \au{Pochelon, A.}, \au{Razumova, K.}, \au{Reimerdes, H.},
  \au{Refke, A.}, \au{Rommers, J.}, \au{Roy, I.}, \au{Sauter, O.}, \au{Suttrop,
  W.}, \au{Toledo, W.~V.}, \au{Tonetti, G.}, \au{Tran, M.~Q.}, \au{Troyon, F.},
  \au{Vyas, P.} \& \au{Ward, D.~J.}} \yr{1999}  \at{{Effect of plasma shape on
  confinement and MHD behaviour in TCV}}.  \jt{Plasma Physics and Controlled
  Fusion}  \bvol{39}~(12B),  \pg{B135--B144}.

\bibitem[Wenninger {\em et~al.\/}(2015)Wenninger, Arbeiter, Aubert,
  Aho-Mantila, Albanese, Ambrosino, Angioni, Artaud, Bernert, Fable, Fasoli,
  Federici, Garcia, Giruzzi, Jenko, Maget, Mattei, Maviglia, Poli, Ramogida,
  Reux, Schneider, Sieglin, Villone, Wischmeier \& Zohm]{Wenninger2015}
{\sc \au{Wenninger, R.}, \au{Arbeiter, F.}, \au{Aubert, J.}, \au{Aho-Mantila,
  L.}, \au{Albanese, R.}, \au{Ambrosino, R.}, \au{Angioni, C.}, \au{Artaud,
  J.-F.}, \au{Bernert, M.}, \au{Fable, E.}, \au{Fasoli, A.}, \au{Federici, G.},
  \au{Garcia, J.}, \au{Giruzzi, G.}, \au{Jenko, F.}, \au{Maget, P.},
  \au{Mattei, M.}, \au{Maviglia, F.}, \au{Poli, E.}, \au{Ramogida, G.},
  \au{Reux, C.}, \au{Schneider, M.}, \au{Sieglin, B.}, \au{Villone, F.},
  \au{Wischmeier, M.} \& \au{Zohm, H.}} \yr{2015}  \at{{Advances in the physics
  basis for the European DEMO design}}.  \jt{Nuclear Fusion}  \bvol{55},
  \pg{063003}.

\bibitem[White {\em et~al.\/}(2013)White, Howard, Greenwald, Reinke, Sung,
  Baek, Barnes, Candy, Dominguez, Ernst, Gao, Hubbard, Hughes, Lin, Mikkelsen,
  Parra, Porkolab, Rice, Walk, Wukitch \& Team]{White2013}
{\sc \au{White, A.~E.}, \au{Howard, N.~T.}, \au{Greenwald, M.}, \au{Reinke,
  M.~L.}, \au{Sung, C.}, \au{Baek, S.}, \au{Barnes, M.}, \au{Candy, J.},
  \au{Dominguez, A.}, \au{Ernst, D.}, \au{Gao, C.}, \au{Hubbard, a.~E.},
  \au{Hughes, J.~W.}, \au{Lin, Y.}, \au{Mikkelsen, D.}, \au{Parra, F.~I.},
  \au{Porkolab, M.}, \au{Rice, J.~E.}, \au{Walk, J.}, \au{Wukitch, S.~J.} \&
  \au{Team, A. C.-M.}} \yr{2013}  \at{{Multi-channel transport experiments at
  Alcator C-Mod and comparison with gyrokinetic simulations}}.  \jt{Physics of
  Plasmas}  \bvol{20}~(5),  \pg{056106}.

\bibitem[van Wyk {\em et~al.\/}(2016)van Wyk, Highcock, Schekochihin, Roach,
  Field \& Dorland]{VanWyk2016}
{\sc \au{van Wyk, F.}, \au{Highcock, E.~G.}, \au{Schekochihin, A.~A.},
  \au{Roach, C.~M.}, \au{Field, A.~R.} \& \au{Dorland, W.}} \yr{2016}
  \at{{Transition to subcritical turbulence in a tokamak plasma}}.  \jt{Journal
  of Plasma Physics}  \bvol{82}~(6),  \arxiv{arXiv: 1607.08173}.

\bibitem[Xanthopoulos {\em et~al.\/}(2014)Xanthopoulos, Mynick, Helander,
  Turkin, Plunk, Jenko, G{\"{o}}rler, Told, Bird \& Proll]{Xanthopoulos2014}
{\sc \au{Xanthopoulos, P.}, \au{Mynick, H.~E.}, \au{Helander, P.}, \au{Turkin,
  Y.}, \au{Plunk, G.~G.}, \au{Jenko, F.}, \au{G{\"{o}}rler, T.}, \au{Told, D.},
  \au{Bird, T.} \& \au{Proll, J.~H.}} \yr{2014}  \at{{Controlling turbulence in
  present and future stellarators}}.  \jt{Physical Review Letters}
  \bvol{113}~(15),  \pg{1--4}.

\bibitem[Xiao {\em et~al.\/}(2007)Xiao, Catto \& Dorland]{Xiao2007}
{\sc \au{Xiao, Y.}, \au{Catto, P.~J.} \& \au{Dorland, W.}} \yr{2007}
  \at{Effects of finite poloidal gyroradius, shaping, and collisions on the
  zonal flow residual}.  \jt{Physics of plasmas}  \bvol{14}~(5),  \pg{055910}.

\bibitem[Zohm {\em et~al.\/}(2013)Zohm, Angioni, Fable, Federici, Gantenbein,
  Hartmann, Lackner, Poli, Porte, Sauter, Tardini, Ward \&
  Wischmeier]{Zohm2013}
{\sc \au{Zohm, H.}, \au{Angioni, C.}, \au{Fable, E.}, \au{Federici, G.},
  \au{Gantenbein, G.}, \au{Hartmann, T.}, \au{Lackner, K.}, \au{Poli, E.},
  \au{Porte, L.}, \au{Sauter, O.}, \au{Tardini, G.}, \au{Ward, D.} \&
  \au{Wischmeier, M.}} \yr{2013}  \at{{On the physics guidelines for a tokamak
  DEMO}}.  \jt{Nuclear Fusion}  \bvol{53}~(7),  \pg{073019}.

\end{thebibliography}

%\getenv{\HOME}{HOME}
%
%Home is \HOME

%\printbibliography
%\end{refsection}

%\bibliography{library}

\end{document}